\documentclass[12pt]{iopart}
\usepackage{iopams}
\usepackage{dsfont}
\usepackage{graphicx}  
\usepackage{color}
\usepackage{hyperref}
\newcommand{\ma}[1]{\begin{eqnarray}{#1}\end{eqnarray}}

\newcommand{\bra}[1]{\langle #1 |}
\newcommand{\ket}[1]{|#1\rangle}

\begin{document}

\title[The quantum transverse-field Ising chain in circuit QED]{The quantum
transverse-field Ising chain in circuit QED: effects of disorder on the
nonequilibrium dynamics}

\author{Oliver Viehmann,$^{1}$ Jan von Delft$^{1}$ and Florian Marquardt$^{2}$}

\address{$^1$ Physics Department,
             Arnold Sommerfeld Center for Theoretical Physics,
             and Center for NanoScience,
             Ludwig-Maximilians-Universit\"at,
             Theresienstra{\ss}e 37,
             80333 M\"unchen, Germany} 
\eads{\mailto{oliver.viehmann@physik.lmu.de}, \mailto{vondelft@lmu.de}}\vspace*{2mm}
\address{$^2$ Institut for Theoretical Physics, Universit\"at Erlangen-N\"urnberg, Staudtstra\ss e 7, 91058 Erlangen, Germany}
\ead{Florian.Marquardt@physik.uni-erlangen.de}
\submitto{\NJP}

\begin{abstract}
We study several dynamical properties of a recently proposed implementation of
the quantum transverse-field Ising chain in the framework of circuit QED. Particular emphasis
is placed on the effects of disorder on the nonequilibrium behavior of the system. 
We show that small amounts of fabrication-induced disorder in the system parameters
do not jeopardize the observation of previously-predicted phenomena. 
Based on a numerical extraction of the mean free path of a wave packet in the system,
we also provide a simple quantitative estimate for certain disorder effects
on the nonequilibrium dynamics of the circuit QED quantum simulator. 
We discuss the transition from weak to strong disorder, characterized by the onset of
Anderson localization of the system's wave functions, and the qualitatively different
dynamics it leads to. 
\end{abstract}


\maketitle

\tableofcontents

\section{Introduction}

Circuit quantum electrodynamics (QED) systems consist of superconducting artificial atoms coupled to the
electromagnetic field in a microwave resonator \cite{Schoelkopf2008}.
Such systems have been successfully
used for implementations of elementary quantum optical Hamiltonians
\cite{Blais2004,Wallraff2004} and basic quantum
information processing \cite{DiCarlo2010,Mariantoni2011,Fedorov2012,Reed2012}.
The rapid technological development in the field of circuit QED will soon facilitate
experiments with highly coherent multi-atom, multi-resonator circuit QED architectures.
This makes circuit QED a promising platform for observing interesting multi-atom quantum
optical effects \cite{Nataf2010a,Delanty2011,Viehmann2011} and even for simulating
genuinely interacting quantum many-body systems from solid state physics
\cite{Romito2005,Hartmann2006,Greentree2006,Wang2007,Koch2009,Tian2010,Schiro2012,Houck2012,Hwang2012,Viehmann2012}.

In \cite{Viehmann2012}, we have proposed and analyzed a circuit QED design that implements the
quantum transverse-field Ising chain (TFIC) coupled to a microwave resonator for
readout. The TFIC is an elementary example of an integrable quantum many-body
system. Despite its simplicity, it still exhibits interesting features, e.g.\
a quantum phase transition (QPT), and therefore serves as a model example system in
the theory of quantum criticality \cite{Sachdev1999} and nonequilibrium
thermodynamics \cite{Polkovnikov2011}.  
Our circuit QED quantum simulator can be used to study quench dynamics, the
propagation of localized excitations, and other nonequilibrium phenomena in the TFIC, 
based on a design that could easily be extended to break the integrability of the
system. While in \cite{Viehmann2012} we have focussed on an idealized 
implementation of the TFIC with perfectly uniform parameters, the main purpose of the
present article is to investigate the effects of disorder in the system parameters on
the dynamical behavior of our quantum simulator. 

The study of disorder effects on quantum simulators is relevant for two reasons.
First, on the more practical level, any real experimental system will come with a
degree of unwanted disorder (especially in condensed matter settings). In the case of
circuit QED systems, inhomogeneities of the system parameters are caused by
fabrication issues as well as by static noise fields (e.g.\ produced by defects).
It is important to verify that the basic behavior of a quantum simulator survives
the amounts of disorder which are present in realistic systems, or even to estimate the amount of disorder that can
be tolerated. Second, on the more fundamental level, simulating quantum many-body
systems with built-in (potentially tunable) disorder is interesting in its own right.
Many physical phenomena, from free propagation of wave packets to quench dynamics to (quantum) phase
transitions can be affected in significant ways by disorder, and this leads to
phenomena such as Anderson localization or disorder-induced phases. 

To prepare our study, we briefly review the system
(\sref{subsec:Implementation}), discuss sources of disorder and
how disorder scales with the tunable system parameters 
(\sref{subsec:Disorder_and_tunability}), 
and explain the mathematical
approach to and some properties of the quantum Ising chain
(\sref{subsec:mathematical_preliminaries}). 
We start our main discussion by considering the time-dependent correlations of the order
parameter of the chain, where the finite-size effects and the long-time behaviour will be
analyzed in the absence of disorder (\sref{subsec:Autocorrelator}). 
Based on this, we will move on to the spectrum of the
resonator coupled to the quantum Ising chain in our system, which is closely related
to the aforementioned time-dependent correlations. To that end, we employ a very
useful approximation which we have introduced in \cite{Viehmann2012} and which will
presumably become important also for future studies of quantum many-body systems
coupled to resonators. In this approximation, the full quantum many-body system is
replaced by a bath of harmonic oscillators with identical spectrum. We show here
that this approximation actually works very well under
appropriate circumstances (\sref{subsec:linear_approximation}). We then
calculate the spectrum of the resonator coupled to a slightly disordered Ising chain
and find that the effects of disorder on the spectrum are small (\sref{subsec:Disorder_Spectrum}). 
The Ising chain in our circuit-QED quantum simulator can be driven out of equilibrium
in several ways. This allows one to perform various types of nonequilibrium
experiments, a particularly appealing application of our setup. In our previous work,
we have suggested to observe the propagation of a localized
excitation through the chain or the nonequilibrium dynamics of the system after a quantum
quench. Here, we show that the predicted phenomena are insensitive to a small amount of
disorder in the system parameters (\sref{subsec:loc_excitation} and 
\sref{subsec:quantum_quenches}, respectively). Moreover, we provide a simple estimate 
for the amount of disorder that will qualitatively change the wave functions and, thus,
strongly affect the dynamics even of small systems (that is, on the scale of neighbouring
artificial atoms). However, as argued above, it would be highly desirable to possess also
a quantitative theory of disorder effects. Since the nonequilibrium dynamics of the uniform TFIC is determined
by the ballistic propagation of quasiparticles (wave packets), we formulate and numerically
verify for the weakly disordered case a relation between the mean
free path of the latter and the parameters of the system and the disorder potential.
By means of this relation we are able to predict the dynamical behavior of our quantum 
simulator given a certain disorder strength, and to estimate the amount of
disorder that a particular experiment can tolerate (\sref{subsec:loc_excitation}).

\section{The quantum transverse-field Ising chain in circuit QED} 

\subsection{Setup} \label{subsec:Implementation}

\begin{figure}
	\centering
	\includegraphics[width=0.7\columnwidth]{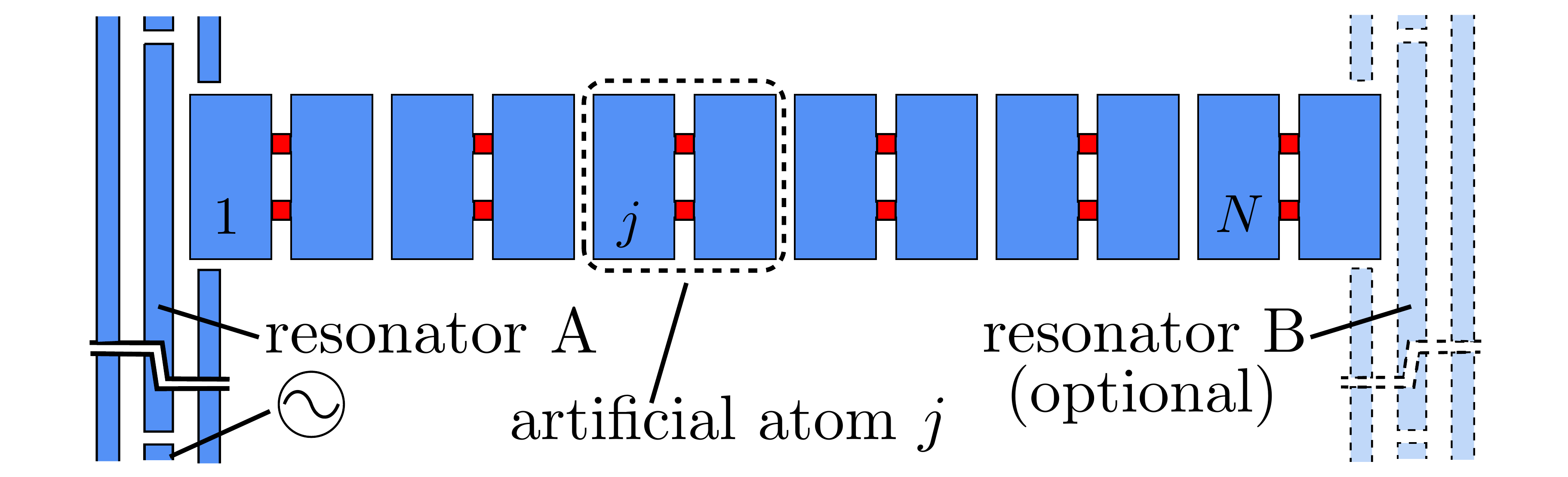}
	\caption{
	Circuit QED implementation of the quantum transverse-field Ising chain (adapted from
    \cite{Viehmann2012}). Charge-based artificial atoms are capacitively coupled to their nearest neighbors. 
    Coupling the first ($N$th) artificial atom to resonator A (B) allows one to use
    standard circuit QED techniques for initialization and read-out of the first
    ($N$th) artificial atom. 
	}
	\label{fig:1}
\end{figure} 
We consider a circuit QED quantum simulator of the TFIC as proposed in
\cite{Viehmann2012}. It consists of a chain of $N$ capacitively coupled charge-based
superconducting artificial atoms \cite{Clarke2008}, such as transmons or Cooper-pair
boxes (the latter have to be biased to their charge degeneracy point
\cite{Clarke2008} to properly simulate the TFIC). For a
review on superconducting artificial atoms, see \cite{Clarke2008}.
The first artificial atom is capacitively coupled to a microwave
resonator (see \Fref{fig:1}). 
This resonator A is required for initialization and readout of the first artificial
atom. For certain types of experiments, e.g., for measuring end-to-end correlators, one
also needs a second resonator B, coupled to the $N$th artificial atom. For details on
the implementation and the theoretical description of the system, see \cite{Viehmann2012}. 
The system (at first, only with resonator A) can be approximately described by the
Hamiltonian
\ma{
\mathcal{H}= \omega_0 a^\dagger a + g (a^\dagger + a) \sigma_x^1 + \mathcal{H}_{\rm
I}, \label{eq:full_system}
}
and $\mathcal{H}_{\rm I}$ is the Hamiltonian of the TFIC,
\begin{eqnarray}
\mathcal{H}_{\rm I} =  \sum_{j=1}^N \frac{\Omega_j}{2} \sigma_z^j - \sum_{j=1}^{N-1}
\mathcal{J}_j \sigma_x^j \sigma_x^{j+1}. \label{eq:TFIC}
\end{eqnarray}
Here, $\sigma_{x/z}^j$ is a Pauli matrix. That is, the artificial atoms are
considered as two-level systems (qubits), and their two states are described as
spin states. The operators $a^\dagger$ and $a$ generate
and annihilate a photon of energy $\omega_0$. The transition frequency $\Omega_j>0$
of the $j$th qubit corresponds to a local
magnetic field acting on the $j$th spin in the usual interpretation of the TFIC. As
such, it would be transverse to the direction of the qubit-qubit coupling
$\mathcal{J}_j $. The latter can be either ferromagnetic
($\mathcal{J}_j>0$, as in the geometry of \Fref{fig:1}) or
anti-ferromagnetic ($\mathcal{J}_j<0$, if the qubits in \Fref{fig:1} are rotated by
$90^\circ$). While in our previous work we have focussed on the uniform
case $\mathcal{J}_j = \mathcal{J}$ and $\Omega_j = \Omega$ for all $j$, we are here often
interested in the case where these system parameters are explicitly nonuniform. 
This is because, on the one hand, a slight nonuniformity of the $\Omega_j$ and $\mathcal{J}_j$
has to be expected from imperfections of the fabrication process. On the other hand,
one can also intentionally detune one or several qubits
(by threading the SQUID-like loops of the qubits with different fluxes)
and observe how the system's properties change depending on the
detuning.

\subsection{Disorder and tunability of the system parameters}
\label{subsec:Disorder_and_tunability}
Let us discuss the flux-tunability and the undesired disorder of the system
parameters in some more detail. We will argue that the qubit transition frequencies
$\Omega_j$ and the qubit-qubit couplings $\mathcal{J}_j$,
when normalized to their respective mean values, may be assumed to be
flux-independent. This will be relevant for our theoretical description of the disorder in the system.

In reality, it should be possible to engineer the geometry of the qubits essentially
uniform. That is, the areas of the qubits' SQUID loops, their charging energies, and
the coupling capacitances between the qubits will only vary weakly in the chain. However, the
(flux-tunable) total Josephson energies $E_\mathrm{J}(\Phi)$
of the artificial atoms should be experimentally harder to control since
these depend exponentially on the properties of the Josephson junctions.
For a flux-tunable (i.e.\ SQUID-type) artificial atom with two Josephson junctions \cite{Koch2007},
\ma{
E_\mathrm{J}(\Phi) = (\epsilon_{\mathrm{J}}^1 + \epsilon_{\mathrm{J}}^2)  \cos \left(
\frac{ \Phi \pi}{ \Phi_0} \right) \left( 1 + d^2 \tan^2\left(\frac{ \Phi \pi}{
\Phi_0} \right) \right)^{1/2}.
}
Here, $\epsilon_{\mathrm{J}}^i$ is the Josephson coupling energy of one Josephson
junction, $\Phi_0$ is the superconducting flux quantum, $\Phi$ is the tunable external flux threading
the SQUID loop, and $d = (\epsilon_{\mathrm{J}}^1 - \epsilon_{\mathrm{J}}^2)/
(\epsilon_{\mathrm{J}}^1 + \epsilon_{\mathrm{J}}^2)$. Assuming equal qubit geometries, $\Phi$ can be chosen identical for all
qubits (e.g.\ by using a common flux line) and only the
$\epsilon_{\mathrm{J}}^i$ can give rise to disorder. 
Even if one allows for $|d| \sim 0.1$, this still means $d^2 \ll
1$, and one can approximate the total Josephson energy of the $j$th artificial atom
 by $E_{\mathrm{J}j}(\Phi) \approx
(\epsilon_{\mathrm{J}j}^1 + \epsilon_{\mathrm{J}j}^2 ) \cos( \Phi \pi / \Phi_0) $ (as
long as $|\Phi| \not\approx \Phi_0/2$).
Now, for Cooper-pair boxes at the charge degeneracy point $\Omega_j(\Phi) \approx
E_{\mathrm{J}j}(\Phi)$, and for transmons $\Omega_j(\Phi) \approx [8
E_{\mathrm{J}j}(\Phi) E_\mathrm{C}]^{1/2}$ \cite{Koch2007}. Thus, under the
assumption of identical geometry, both for Cooper-pair boxes and transmons
the transition frequencies $\Omega_j(\Phi)$ of all qubits $j$ scale with a $j$-independent
function $\alpha(\Phi)$ of the (global) flux $\Phi$, $\Omega_j(\Phi) =
\alpha(\Phi) \Omega_j(0)$. Here, $\alpha(\Phi) = \cos (\Phi
\pi/ \Phi_0)$ for Cooper-pair boxes and $\alpha(\Phi) =[\cos( \Phi \pi/ \Phi_0)]^{1/2}$ for transmons.
This result implies that the qubit transition frequencies, 
when normalized to their flux-dependent mean value, do not depend on 
$\Phi$ and, hence, have the same statistical properties for all $\Phi$. 
Explicitly, the mean value of the $\Omega_j$ is given by
$\overline{\Omega_j(\Phi)} = \alpha(\Phi) \overline{\Omega_j (0)}$.
Thus, the mean value of the $\Omega_j$ is flux tunable. However, the 
normalized qubit transition frequencies $\Omega_j(\Phi)/\overline{\Omega_j(\Phi)}$ are
independent of $\Phi$, which must also be the case, for instance, for their standard
deviation. This will become important for our numerical
implementation of disorder in the $\Omega_j$ when
we consider changes of the external magnetic flux $\Phi$.

The qubit-qubit couplings $\mathcal{J}_j$ can also depend on the $E_{\mathrm{J}j}(\Phi)$
and, thus, on the $\Omega_j$. This is the case for transmons, where approximately
$\mathcal{J}_j \propto (\Omega_j \Omega_{j+1})^{1/2} \propto (E_{\mathrm{J}j}
E_{\mathrm{J}j+1})^{1/4}$ \cite{Dewes2012,Viehmann2012}. That is, the disorder in the
$\Omega_j$ and the $\mathcal{J}_j$ will not be independent for transmons. Moreover,
$\Omega_j$, $\mathcal{J}_j$, and their mean values $\overline{\Omega_j}$ and
$\overline{\mathcal{J}_j}$ 
change with the external flux $\Phi$ approximately in the same proportion $(\propto[ \cos(\Phi
\pi/\Phi_0)]^{1/2})$. For Cooper-pair boxes, on the other hand,
the $\mathcal{J}_j$ depend only on charging energies and
not on the $E_{\mathrm{J}j}(\Phi)$ \cite{Viehmann2012}. This means that 
the $\mathcal{J}_j$ are not affected by changes of the external flux. Furthermore,
the disorder in the
$\mathcal{J}_j$ should be less pronounced than and hardly correlated with the 
disorder in the $\Omega_j$. Concerning the relative strength and the correlation of the disorder in the
$\mathcal{J}_j$ and the $\Omega_j$, we remark that also static noise fields can play
a role, producing some disorder also in the various charging energies of the system (in
particular for Cooper-pair boxes, which have small electrostatic capacitances). Apart
from that, disorder in the $\mathcal{J}_j$ will turn out to have a
much weaker effect than disorder in the $\Omega_j$. These deliberations justify to
assume for simplicity that both for Cooper-pair boxes and for transmons disorder in
the $\Omega_j$ and $\mathcal{J}_j$ can be present to a comparable degree, 
and that disorder in the $\Omega_j$ ($\mathcal{J}_j$) would be 
uncorrelated with the disorder possibly present in the $\mathcal{J}_j$ ($\Omega_j$).
We finally remark that many properties of the transverse-field Ising chain are
determined by the ratio $\overline{\Omega_j}/\overline{\mathcal{J}_j}$, since
this ratio essentially (in the limit of weak disorder) determines the eigenstates of
the system (see below). For standard transmons, the ratio $\overline{\Omega_j}/\overline{\mathcal{J}_j}$ is not straightforwardly
flux-tunable. One of the experiments we suggest to do with our quantum simulator
relies on the possibility to change the eigenfunctions of the system [cf.
\sref{subsec:quantum_quenches}], which can be performed only by changing the ratio
$\overline{\Omega_j}/\overline{\mathcal{J}_j}$. All other possible experiments discussed in this
article can be done in principle with Cooper-pair boxes and transmons equally well, 
irrespective of the $\mathcal{J}_j$ being flux-dependent or not \cite{Viehmann2012}. 
Therefore, when plotting our results as function of a flux-tunable system parameter, we will
assume for definiteness that our circuit QED quantum simulator of the TFIC
is implemented with Cooper-pair boxes, and that the $\mathcal{J}_j$ do
not change with the external magnetic flux.

\subsection{The transverse-field Ising chain } \label{subsec:mathematical_preliminaries}
The Hamiltonian \eref{eq:TFIC} can be exactly diagonalized by means of a
Jordan-Wigner transformation, which was first used in this context in
\cite{Lieb1961,Pfeuty1970}. This transformation maps the spin degrees of freedom to
fermionic operators $c_j,c_j^\dagger$ via $\sigma_j^+ = c_j^\dagger \exp(i \pi
\sum_{k=1}^{j-1} c_k^\dagger c_k )$ and yields
\ma{
\mathcal{H}_I = - \sum_{j=1}^N \frac{\Omega_j}{2} +  \sum_{j=1}^N \Omega_j c_j^\dagger c_j - \sum_{j=1}^{N-1} \mathcal{J}_j[ c_j^\dagger c_{j+1}^\dagger +
c_j^\dagger c_{j+1} + \mathrm{H.c.}] ,\label{eq:cfermions}
}
Up to a constant $- \sum_{j} \Omega_j/2$, this Hamiltonian is of the form 
\begin{eqnarray}
H = \sum_{i,j=1}^N [ c_i^\dagger A_{i,j} c_j + 1/2 (c_i^\dagger B_{i,j} c_j^\dagger + \mathrm{H.c.})].  \label{eq:quadratic_Hamiltonian}
\end{eqnarray}
Note that the conditions $H=H^\dagger$ and $\{c_j,c_j^\dagger\}=1$ require
$A=A^\dagger$ and $B=-B^T$. By introducing new fermions $\eta_k = \sum_{j=1}^N
g_{k,j} c_j + h_{k,j} c_j^\dagger$, such Hamiltonians can be transformed into the
diagonal form
$H=\sum_k \Lambda_k (\eta_k^\dagger \eta_k -1/2) + \sum_j A_{j,j}/2$ \cite{Lieb1961}.
The components $g_{k,j}$ and $h_{k,j}$ of the vectors $g_k$ and $h_k$ and the
excitation energies $\Lambda_k$ of $H$ are determined by defining normalized vectors
$\phi_k = g_k + h_k$ and $\psi_k = g_k - h_k$ and by solving the equations 
\begin{eqnarray}
\phi_k(A-B) = \Lambda_k \psi_k,\qquad
\psi_k(A+B) = \Lambda_k \phi_k. \label{eq:relpsiphi}
\end{eqnarray}
In our case, 
\begin{eqnarray}
A = 
\left(
   \begin{array}{cccccc}
     \Omega_1 & -\mathcal{J}_1 & 0 &  & \cdots & 0\\
     -\mathcal{J}_1 & \Omega_2 & -\mathcal{J}_2 & & & \\
     0 & -\mathcal{J}_2 & \Omega_3 & -\mathcal{J}_3 & & \\
     \vdots & & \ddots & \ddots & \ddots & \\ 
      & & & -\mathcal{J}_{N-2}&\Omega_{N-1} &-\mathcal{J}_{N-1}\\
	0 & & & & -\mathcal{J}_{N-1} &\Omega_N   
   \end{array}
\right), \label{eq:MatrixA}
\end{eqnarray}
and $B$ is obtained by substituting $A_{j,j} = \Omega_j \rightarrow 0 $ and
$A_{j+1,j}=-\mathcal{J}_j \rightarrow \mathcal{J}_j$ in $A$. For uniform $\Omega_j$
and $\mathcal{J}_j$,  the $\phi_k$, $\psi_k$, and $\Lambda_k$ can be analytically
calculated from Equations \eref{eq:relpsiphi} (see, e.g., \cite{Viehmann2012}). For
nonuniform system parameters, these quantities have to be determined numerically. In
both cases, the Hamiltonian $\mathcal{H}_{\rm I}$ of the TFIC can be written in the
form
\ma{
\mathcal{H}_{\rm I} = \sum_k \Lambda_k (\eta_k^\dagger \eta_k -1/2),
}
and knowledge of the $\phi_k$ and $\psi_k$ allows one to express spin observables in
terms of the $\eta_k$-fermions, which is the basis of many of our calculations. For
instance,
\ma{
\sigma_z^j = (c_j^\dagger + c_j)(c_j - c_j^\dagger) = \sum_{k,k^\prime} \phi_{k,j}
\psi_{k^\prime,j} (\eta_k^\dagger  + \eta_k) (\eta_{k^\prime} -
\eta_{k^\prime}^\dagger).
}
We collect some important facts about the TFIC. In the uniform case, 
\ma{
\Lambda_k=2J \sqrt{1+\xi^2 -2\xi \cos k}.\label{eq:eigenvaluesTFIC}
}
Here, $J=|\mathcal{J}|$ and $\xi=\Omega/2\mathcal{J}$ is the normalized transverse
field. The possible values of $k$ are solutions of $\sin k N= \xi \sin k (N+1)$. For
$N \rightarrow \infty$, the uniform TFIC undergoes a second order QPT at $\xi = \pm
1$ from a ferromagnetic [$\xi \in (0,1)$] or an anti-ferromagnetic [$\xi \in (-1,0)$] 
ordered phase with doubly
degenerate eigenstates (one $\Lambda_k \rightarrow 0 $) to a paramagnetic disordered
phase with $\Lambda_k > 0$ for all $k$. The QPT is signalled by the disappearance of
long-range correlations in $\sigma_x$. This QPT will also occur in a nonuniform
system (at some mean transverse field strength $\overline{\Omega_j}$)
\cite{Sachdev1999}. However, there can be weakly (dis)ordered Griffith-McCoy
`phases' in the vicinity of the critical point
\cite{Griffiths1969,McCoy1969,Fisher1992,Fisher1995}.

Finally, we introduce a convenient notation for nonuniform $\Omega_j$ and
$\mathcal{J}_j$. In this case, we will frequently write $\Omega_j = \Omega \tau_j$
and $\mathcal{J}_j = \mathcal{J} \tau^\prime_j$, where $\tau_j$ and $\tau^\prime_j$
usually have mean $1$, or, if $\Omega_j$ and $\mathcal{J}_j$ follow probability
distributions, expectation value $1$. We will refer to $\Omega$ as the `mean' qubit
transition frequency, even if $\Omega = \langle \Omega_j \rangle$ is the expectation
value of a probability distribution and the actual mean value $\overline{\Omega_j}$
is (for finite $N$) in general different from $\Omega$. We use the same convention
for the qubit-qubit coupling $\mathcal{J}$. Furthermore, we define the local
and the `mean' normalized transverse magnetic field, $\xi_j = \Omega_j/2\mathcal{J}_j$ and
$\xi = \Omega/ 2\mathcal{J}$. Note that in general both $\xi \neq \overline{\xi_j}$
and $\xi \neq \langle \xi_j\rangle$ (but for the probability distributions we will
consider, $(\xi- \langle \xi_j \rangle)/\langle \xi_j \rangle <1\%$).
We usually characterize $\mathcal{H}_{\rm I}$ by the parameters 
$\xi$, $J$, $\tau_j$, and $\tau^\prime_j$.
Under the assumptions formulated in \sref{subsec:Implementation}, $\Omega$ and thus
$\xi$ are flux-tunable without changing the $\tau_j$ in the proposed circuit QED quantum
simulator of the TFIC.

\section{Spectrum of the system}

In order to provide a guideline for the initial experimental characterization of
our setup, we have calculated in \cite{Viehmann2012}
the transmission spectrum $S$ of the resonator as a function of the probe
frequency $\omega$ and the flux-tunable qubit transition frequency $\Omega$ [see
below, equation \eref{eq:Scoupled}].
To that end, we have first calculated the spectrum of the bare TFIC
for coupling to the first qubit via $\sigma_x^1$,
\ma{
\tilde{\rho}(\omega)=\int \mathrm{d}t \rme^{\rmi \omega t} \langle \sigma_x^1(t) \sigma_x^1(0) \rangle,
}
which is the Fourier transform of the qubit autocorrelator $\rho(t)=\langle
\sigma_x^1(t) \sigma_x^1(0) \rangle $. We have argued that for sufficiently large
(but finite) $N$, qubit decay processes will render the measured spectrum continuous
and akin to the spectrum one would obtain by taking the limit $N \rightarrow \infty$
in the calculation of $\rho$. Assuming small coupling $g/\omega_0 \ll 1$ of
the first qubit and the resonator, we have then considered the TFIC as a linear bath
for the resonator, and this approximation allowed us to calculate the resonator
spectrum $S$ in the coupled system. In this section, we add some remarks on the
interpretation of the autocorrelator, the transition $N\rightarrow \infty$, and the linear approximation. 
Moreover, we discuss how a small amount of disorder in the qubit parameters 
due to imperfections in the fabrication process affects the resonator spectrum~$S$.

\subsection{Time-dependent correlations in the transverse-field Ising chain} \label{subsec:Autocorrelator}
\begin{figure}
	\centering
	\includegraphics[width=0.6\columnwidth]{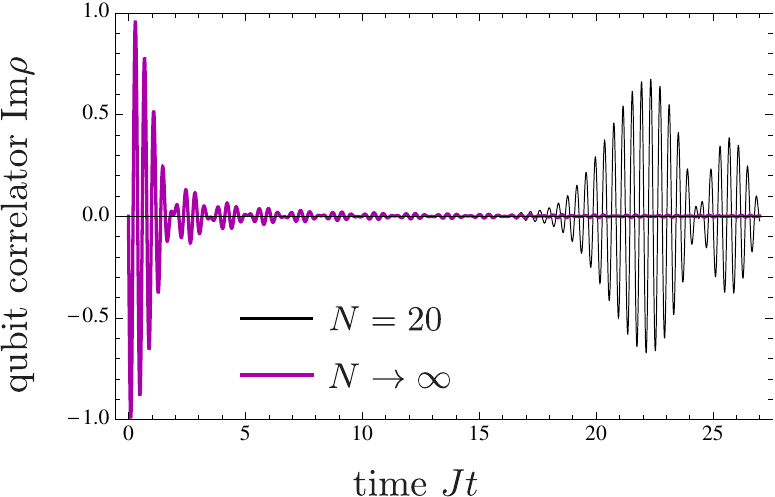} 
	\caption{
	Imaginary part of the qubit autocorrelator $\rho(t)= \langle \sigma_x^1(t)
    \sigma_x^1(0) \rangle$ of the transverse-field Ising chain
	with normalized transverse field $\xi= \Omega / 2 \mathcal{J} = 8$ in the cases $N=20$ (black) 
	and $N\rightarrow \infty$ (magenta).  
}
	\label{fig:2}
\end{figure} 
By means of the spin--free-fermion mapping described in \sref{subsec:mathematical_preliminaries}, one readily finds
\begin{eqnarray}
 \rho(t) = \langle \sigma_x^1(t) \sigma_x^1(0) \rangle =  \sum_{k} \phi_{k,1}^2
 \rme^{-\rmi t \Lambda_k}.
\label{eq:autocorrelator}
\end{eqnarray}
Here and in the following, expectation values are calculated under the 
assumption of zero temperature. This is justified because the band gap of the Ising
chain is of the same order of magnitude as the qubit transition frequencies $\Omega \sim 5$ GHz (except near
the critical point) and, thus, much bigger than the usual mK temperatures of a cryogenic
environment.
In the uniform case $\Omega_j = \Omega $
and $\mathcal{J}_j = \mathcal{J}$, where explicit expressions for $\phi_k$ and
$\Lambda_k$ can be found, the limit $N \rightarrow \infty$ can be taken analytically
and yields \cite{Viehmann2012}
\begin{eqnarray}
\rho(t) = \Theta  (1-|\xi|) (1-|\xi|^2)+ \frac{2 }{\pi} \int_0^\pi \mathrm{d} k
\frac{ \xi^2 \sin^2 k}{1+\xi^2 -2\xi \cos k} \rme^{-\rmi t \Lambda(k)}.
\label{eq:autocorrcont}
\end{eqnarray}
Here, $\Theta(x)$ is the Heaviside step function and $\Lambda(k)$ stands for
$\Lambda_k$ with continuous $k$ [equation \eref{eq:eigenvaluesTFIC}].
The first term on the RHS of \eref{eq:autocorrcont}
causes a nonzero mean value of $\mathrm{Re}\rho(t)$ in the ordered phase.
\Fref{fig:2} shows $\mbox{Im}\rho(t)$ for $\xi=8$ in the cases $N=20$
[equation \eref{eq:autocorrelator}] and $N\rightarrow \infty$ [equation
\eref{eq:autocorrcont}] (the
time evolutions of $\mbox{Re}\rho$ and $\mbox{Im}\rho$ are qualitatively similar and
agree for $|\xi| \gg 1$ up to a phase). For small times, the curves coincide (the
second covers the first). However, the finite size of the TFIC with $N=20$ causes a
revival of $\rho$ at $T_r \approx 2 N/v$ with $v=\max
[\mathrm{d}\Lambda(k)/\mathrm{d}k]$ ($v=2 J |\xi|$ for $\xi<1$ and $v=2 J $ for
$|\xi|>1$). This can be understood in the following way. The autocorrelator $\rho$ is
related to the linear response $\Delta \langle \sigma_x^1 \rangle (t)$ of the TFIC to
a perturbation $ \propto\delta(t) \sigma_x^1 $ relative to the equilibrium value
$\langle \sigma_x^1 \rangle = 0 $. Indeed, Kubo's formula predicts $\Delta \langle
\sigma_x^1 \rangle (t) \propto \mbox{Im} \rho(t)$. The $\delta$-pulse at $t=0$ forces
the first spin in the $-x$-direction. This local excitation in position space is
composed of many excitations in $k$-space. Since most of them have velocity $v$
\cite{Viehmann2012}, the local excitation propagates with velocity $v$ through the
system, is reflected at the far end of the chain, and causes revivals of $\rho$ at
multiples of $T_r=2N/v$. To further clarify the transition $N \rightarrow \infty$,
we note that for large $t$, $\rho$ has a standard deviation from its mean $\propto
1/\sqrt{N}$. This can be expected from \eref{eq:autocorrelator} since $|\rho(t)|^2
\sim 1/N^2 \sum_{k,k^\prime} \rme^{\rmi t(\Lambda_k -\Lambda_{k^\prime})}$ and, for
$t\rightarrow \infty$, all terms in the sum except for those with $k= k^\prime$ will
cancel. In general, the $ t \rightarrow \infty$ fluctuations that we find for all
time-dependent observables considered in this work are due to the finite system size
and decrease with $N$ (but not all of them behave as $\propto 1/\sqrt{N}$).

\subsection{Spectrum of the resonator -- the linear approximation}
\label{subsec:linear_approximation}
Taking the Fourier transform of equations \eref{eq:autocorrelator} and \eref{eq:autocorrcont} yields 
the spectrum $\tilde{\rho}(\omega)$ of the TFIC for a force that couples to $\sigma_x^1$ for finite 
$N$ and $N \rightarrow \infty$, respectively. In order to calculate the spectrum $S$ of
the resonator, whose coordinate $(a^\dagger + a)$ couples to $\sigma_x^1$ [cf.\
equation \eref{eq:full_system}], we have suggested \cite{Viehmann2012} a useful
approximation: We consider the TFIC as a linear bath for the resonator. That is, we replace the
TFIC by a set harmonic oscillators having the spectrum $\tilde{\rho}$ of the TFIC.
This approximation can be straightforwardly generalized to other
contexts, where a different many-body system couples to a resonator. It is
justified in the limit of small qubit-resonator coupling
$g/\omega_0 \ll 1$, as we discuss in the following. 

The linear approximation for the TFIC-bath fails as soon as probing 
the resonator sufficiently excites the TFIC so that its nonlinearity becomes important.
Thus, the linear approximation requires small coupling $g$ and is worst if the TFIC
is on resonance with the resonator ($\omega_0$ within the band $\Lambda_k$ of the TFIC). The "most nonlinear"
bath possible for the resonator, that is, the bath whose nonlinearity becomes
important for the smallest value of $g$, is
a bath consisting of only a single qubit on resonance with the resonator. If the
linear approximation is adequate for such a system in the limit $g/\omega_0 \ll 1$,
it will be also sufficient for our purposes. Therefore, we now consider the case
$N=1$ and $\Omega= \omega_0$ of equation 
\eref{eq:full_system} and calculate the spectrum of the resonator by linearizing the
single-qubit-bath. Since the atomic Hilbert space is small for $N=1$, we can then
numerically check the accuracy of our approximation. We also compare our
approximation with the resonator spectrum calculated analytically within the 
rotating wave approximation (RWA), which is the standard approximation of
$\mathcal{H}$ in this specific situation.

For $N=1$ and $\Omega = \omega_0$, the Hamiltonian $\mathcal{H}$ [equation \eref{eq:full_system}] becomes
\ma{
\mathcal{H}_{N=1}= \omega_0 a^\dagger a + g( a^\dagger + a )\sigma_x +
\frac{\omega_0}{2} \sigma_z
\label{eq:nequone}
}
That is, the resonator coordinate $(a^\dagger + a)$ couples to a single-qubit-bath
with Hamiltonian $\mathcal{H}_{\rm q} = \frac{\omega_0}{2} \sigma_z $ via a force $g \sigma_x$. The
spectrum of this force is 
\ma{
\tilde{F}_{\rm q}(\omega) =  \int \mathrm{d} t \rme^{\rmi \omega t} \; _{\rm
q}\langle g\sigma_x(t)\,g \sigma_x(0) \rangle_{\rm q}
= 2 \pi g^2 \delta (\omega -\omega_0),
}
where the time evolution of $\sigma_x$ and the expectation value $_{\rm q} \langle \,
. \, \rangle_{\rm q}$ are to be calculated with respect to (the ground state of) $\mathcal{H}_{\rm q}$. 
Now we linearize the system and replace $\mathcal{H}_{N=1}$ by 
\ma{
\mathcal{H}_{\rm lin}= \omega_0 a^\dagger a + g^\prime (a^\dagger + a) (b^\dagger +
b) + w b^\dagger b\label{eq:Hlin} 
}
with bosonic $b,b^\dagger$ and parameters $g^\prime$ and $w$ to be determined. In
\eref{eq:Hlin}, the resonator couples to a force $g^\prime (b^\dagger + b)$ exerted
by a bath that consists of a single harmonic oscillator with Hamiltonian
$\mathcal{H}_{\rm ho} = w b^\dagger b$. The spectrum of this force reads
\ma{
\tilde{F}_{\rm ho}(\omega)=  2 \pi (g^\prime)^2\delta(\omega- w).
}
Thus, we choose $g^\prime = g$ and $w= \omega_0$ such that $\tilde{F}_{\rm ho}=
\tilde{F}_{\rm q}$. With this substitution, we now calculate the autocorrelator of
the resonator coordinate, 
\ma{
\rho_{\rm lin} (t) = \, _{\rm lin}\langle [a^\dagger (t) + a
(t) ] [a^\dagger (0) + a (0)] \rangle_{\rm lin} ,
} 
and its Fourier transform, the resonator spectrum
\ma{
\tilde{\rho}_{\rm lin} (\omega) = \int \mathrm{d} t \rme^{\rmi \omega t} \rho_{\rm
lin} (t),
} 
according to \Eref{eq:Hlin}. To that end, we express the resonator coordinate
$(a^\dagger + a)$ in terms of the (bosonic) eigenmodes $\tilde{c}_\pm$ with frequencies 
$\tilde{\omega}_\pm = \sqrt{ \omega_0^2 \pm 2 g \omega_0}$ of $\mathcal{H}_{\rm lin}$,
\ma{
(a^\dagger + a) = \sqrt{\frac{\omega_0}{2}} \left( \frac{\tilde{c}_+^\dagger
+\tilde{c}_+}{\sqrt{\tilde{\omega}_+}} + \frac{\tilde{c}_-^\dagger +
\tilde{c}_-}{ \sqrt{\tilde{\omega}_-}} \right).\label{eq:Bog}
}
Using \eref{eq:Bog}, one readily finds
\begin{eqnarray}
\rho_{\rm lin}(t) = \frac{\omega_0}{2} \left[ 
\frac{\rme^{- \rmi\tilde{\omega}_+ t}}{\tilde{\omega}_+} +
\frac{\rme^{- \rmi\tilde{\omega}_- t}}{\tilde{\omega}_-} \right],
\label{eq:autocorrLin}\\
\tilde{\rho}_{\rm lin} (\omega) = \pi \left[ \frac{\omega_0}{\tilde{\omega}_+}
\delta(\omega - \tilde{\omega}_+) + \frac{\omega_0}{\tilde{\omega}_-} \delta ( \omega
- \tilde{\omega}_-) \right].\label{eq:specLin}
\end{eqnarray}

Before we go on and compare these approximate analytical results with numerical
finite-size calculations for $\mathcal{H}_{N=1}$ [\Eref{eq:nequone}], we calculate the same quantities on
the basis of the standard approximation to $\mathcal{H}_{N=1}$ for $g/\omega_0 \ll
1$, the RWA (see, e.g., \cite{Walls2008}). This will be a helpful
benchmark for estimating the quality of the linear approximation. In the
RWA, the Hamiltonian $\mathcal{H}_{N=1}$
reduces to the Jaynes-Cummings Hamiltonian
\ma{
\mathcal{H}_{\rm RWA} = \omega_0 a^\dagger a + g( a^\dagger \sigma^{-} + a \sigma^{+}) +
\frac{\omega_0}{2} \sigma_z.
\label{eq:Jaynes_Cummings}
}
This Hamiltonian can be straightforwardly diagonalized, and one can therefore
analytically calculate the autocorrelator $\rho_{\rm RWA} (t) $ and the spectrum
$\tilde{\rho}_{\rm RWA} (\omega)$
of the resonator in the approximation provided by $\mathcal{H}_{\rm RWA}$,
\begin{eqnarray}
\rho_{\rm RWA} (t) = \frac{1}{2} \left[ \rme^{ - \rmi t ( \omega_0 + g)} + \rme^{ -
\rmi t (\omega_0 - g)} \right ],\label{eq:autocorrJC} \\ 
\tilde{\rho}_{\rm RWA} (\omega) = \pi [ \delta ( \omega - (\omega_0 + g)) + \delta ( \omega - (\omega_0 -
g))].\label{eq:specJC}
\end{eqnarray}
On the
basis of \eref{eq:Jaynes_Cummings}, the results \eref{eq:autocorrJC} and
\eref{eq:specJC} are exact.

The autocorrelator and the spectrum of the resonator can also be calculated numerically 
after truncating the photonic Hilbert space. This is achieved by
expanding $\mathcal{H}_{N=1}$ and the resonator coordinate $(a^\dagger + a)$ in the
product basis $\{ \ket{s_z,\nu}\}$, where $s_z = \uparrow, \downarrow$ and $\nu \in
\mathds{N}_0$, and dropping all matrix elements with $\nu > \nu_{\rm max}$. In this
finite-size approximation, the eigenvalues $E_n$ and eigenvectors $\ket{n}$ of $\mathcal{H}_{N=1}$ can be
numerically calculated ($n=0, \ldots, n_{\rm max} = 2\nu_{\rm max} + 1$) and give
$\rho(t)$ and $\tilde{\rho} (\omega)$ according to 
\begin{eqnarray}
\rho(t)& = \sum_{n=0}^{n_{\max}} \rme^{ - \rmi (E_n-E_0)t } \,| \bra{0} ( a^\dagger +
a) \ket{n} |^2 \label{eq:autocorrF}\\
\tilde{\rho}(\omega)& =2 \pi \sum_{n=0}^{n_{\max}}    \delta(\omega -(E_n-E_0))\, |
\bra{0} ( a^\dagger + a) \ket{n} |^2 \label{eq:specF}
\end{eqnarray}
Even for relatively strong coupling $g/\omega_0 = 0.3$, the numerical results for
$\rho(t)$ and $\tilde{\rho} (\omega)$ are already converged if $\nu_{\max} = 3$ photonic 
excitations are taken into account. However, to be on the safe side, we choose $\nu_{\max} = 10$ in our
calculations, which is still numerically easily tractable.

\begin{figure}
	\centering
	\includegraphics[width=\columnwidth]{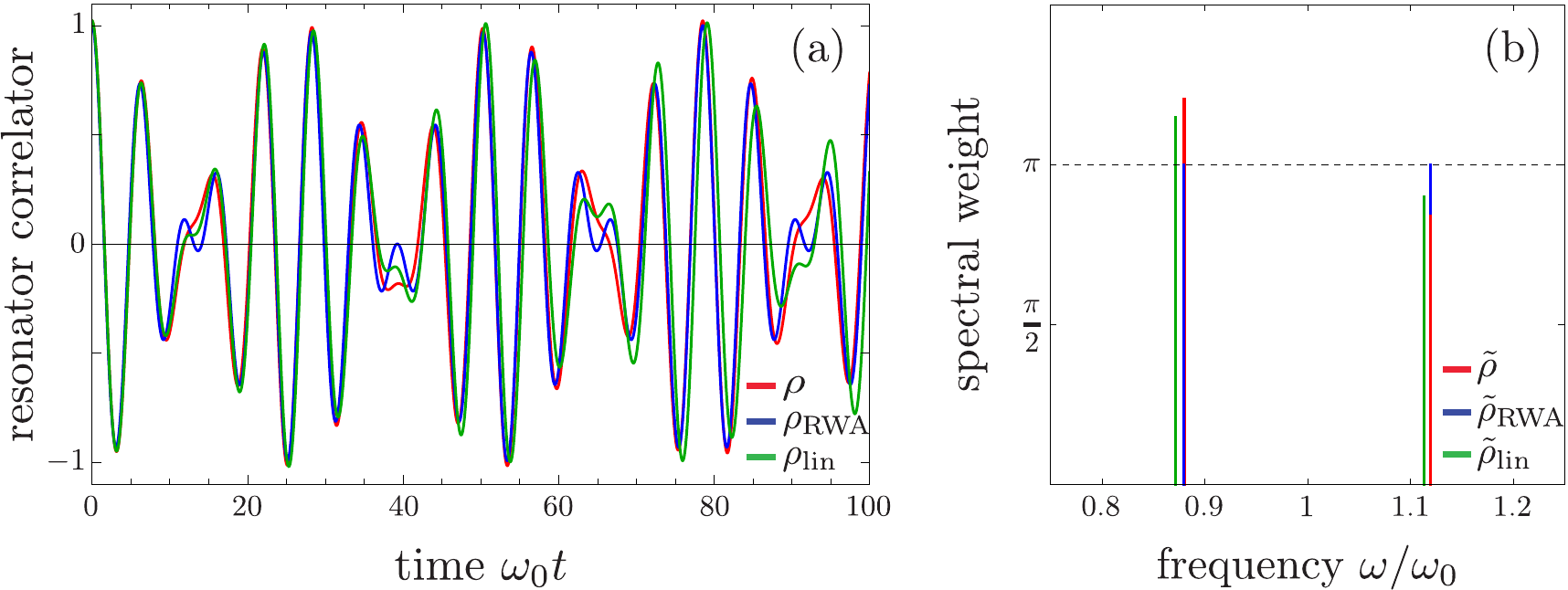} 
	\caption{Comparison of the rotating-wave approximation and the linear
    approximation with highly accurate finite-size numerics for a resonator with
    frequency $\omega_0$ resonantly coupled to a single qubit with coupling strength
    $g/\omega_0=0.12$. (a)
    Autocorrelator $\rho(t)= \langle [ a^\dagger (t) + a(t)] [a^\dagger (0) + a (0)]
    \rangle $ of the resonator (red), and the same quantity calculated within the rotating-wave approximation
    ($\rho_{\rm RWA}$, blue) and the linear approximation ($\rho_{\rm lin}$, green). (b)
    Spectrum $\tilde{\rho}(\omega) = \int \mathrm{d} t e^{\rmi \omega t} \rho(t)$ of
    the resonator (red), and the same quantity calculated within the rotating-wave approximation
    ($\tilde{\rho}_{\rm RWA}$, blue) and the linear approximation ($\tilde{\rho}_{\rm
    lin}$, green). The dashed line is a guide to the eye. 
}
	\label{fig:3}
\end{figure} 
Our results for the autocorrelator $\rho (t)$ and the spectrum $\tilde{\rho}(\omega)$
of the resonator in $\mathcal{H}_{N=1}$ are plotted, respectively, in \fref{fig:3}(a) [equations
\eref{eq:autocorrLin},\eref{eq:autocorrJC},\eref{eq:autocorrF}] and \fref{fig:3}(b) [equations
\eref{eq:specLin},\eref{eq:specJC},\eref{eq:specF}]. In both plots, we choose $g/\omega_0=0.12$, which is
the largest ratio of $g/\omega_0$ used in this work and in \cite{Viehmann2012}. The
autocorrelator $\rho(t) $ of the resonator (red) is well approximated both by
the RWA ($\rho_{\rm RWA}$, blue) and the linear approximation ($\rho_{\rm
lin}$, green), and the quality of these approximations is essentially equal. For small $t$, 
the linear approximation might be even more accurate than the RWA, but becomes worse
at large $t$. This can be understood in the frequency domain. In \fref{fig:3}(b), we
plot the spectral weights of the delta-peaks in the spectra $\tilde{\rho}$,
$\tilde{\rho}_{\rm RWA}$, and $\tilde{\rho}_{\rm lin}$ (red, blue, green) at the corresponding peak
positions. The spectrum $\tilde{\rho}$ contains also delta-peaks at higher
frequencies than the ones plotted, but their weight is virtually zero 
($ 2 \pi | \bra{0} (a^\dagger + a) \ket{n}|^2 < 10^{-7}$ for all
$n \neq 1,2$). Both approximations yield good predictions for the positions and the
spectral weights of the peaks in $\tilde{\rho}$. The RWA is more precise in
predicting the peak positions and the linear approximation in predicting the spectral
weights (note, however, that the peak positions in $\tilde{\rho}_{\rm lin}$ and
$\tilde{\rho}_{\rm RWA}$ agree up to first order in $g/\omega_0$). 
Thus, the linear approximation is more precise for small $t$, in particular 
at $t \approx 0 $ and where the envelope of $\rho(t)$ has a minimum, but becomes
worse for large $t$. In summary, we conclude that even for the situation $N=1$ and
$\Omega=\omega_0$, the linear approximation yields good
results for the autocorrelator and the spectrum of the resonator  in the limit
$g/\omega_0 \ll 1$ that are qualitatively comparable to the usual RWA in this
context. This implies that the linear
approximation is well-justified in our calculation of the spectrum of a resonator
coupled to a TFIC.

\subsection{Spectrum of the resonator -- disorder effects}
\label{subsec:Disorder_Spectrum}
The linear approximation for the TFIC allows one to express the spectrum $S (\omega)$ of the
(coupled) resonator as a function of the spectrum $\tilde{\rho}(\omega)$ of the
TFIC \cite{Viehmann2012},
\ma{
S (\omega) =  \frac{4 \Theta (\omega)  [\kappa + g^2  \tilde{\rho}(\omega)
]}{[\omega^2/\omega_0 - \omega_0 - 4g^2 \chi(\omega^2)]^2 + [\kappa  + g^2
\tilde{\rho}(\omega) ]^2}. \label{eq:Scoupled}
}
Here, $ \kappa $ is the full linewidth at half maximum of the Lorentzian spectrum of
the uncoupled ($g=0$) resonator and $\chi(\omega^2)$ denotes the principal-value
integral $\chi(\omega^2) = 1/(2\pi) \int \mathrm{d} \Omega \tilde{\rho}(\Omega)
\Omega/(\omega^2-\Omega^2)$. This result is actually general and holds for any
linear bath coupled to a resonator, with an arbitrary spectrum $\tilde{\rho}$. Plots
of $S$, with $\tilde{\rho}(\omega)$ being the Fourier transform of
\eref{eq:autocorrcont}, are presented in \cite{Viehmann2012}. However, in an actual
implementation of the proposed setup,
the qubit parameters $\mathcal{J}_j$ and $\Omega_j$ will not be perfectly uniform, due to imperfections in the
fabrication process. We now investigate how this modifies the characteristic features of the spectrum $S$
of the uniform system. It is known in the field of
random-matrix theory that disorder would have to be very strong in order to have a
dominant effect on (average) spectra. We will observe the same here, in this concrete
model system.

\begin{figure}
	\centering
	\includegraphics[width=0.7\columnwidth]{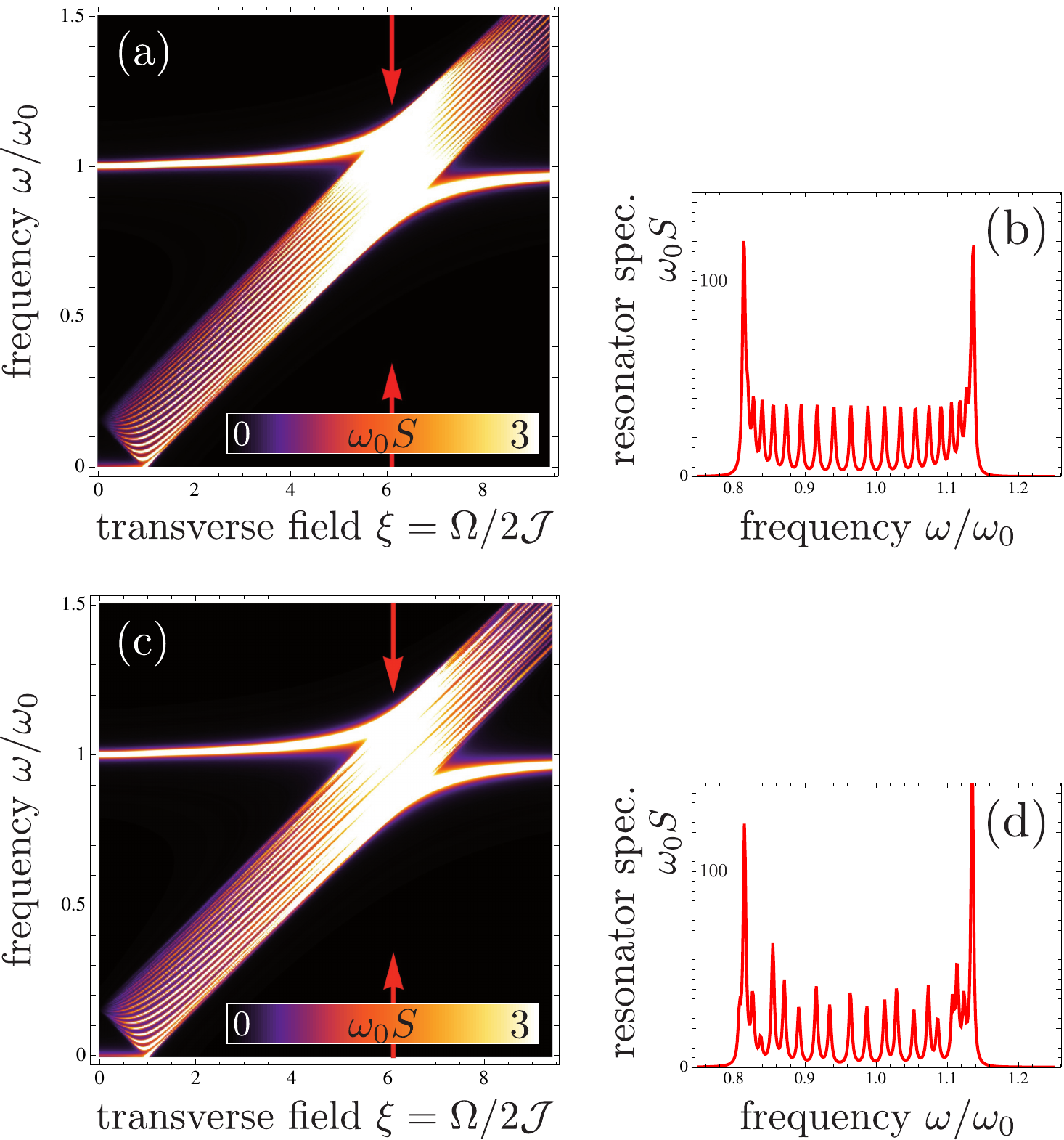}
	\caption{
	(a) Spectrum $ S $ of a resonator coupled to a finite uniform TFIC with $N=20$
	vs.\ probe frequency $\omega$ and normalized transverse field $\xi= \Omega/2\mathcal{J}$.
	The parameters are $g=0.12$, $J=0.08$, $ \kappa = 10^{-4}$, and $ \gamma = 5
    \times 10^{-3}$ (in units of $\omega_0$). For better visibility of the features,
    values $>3$ are plotted white. 
	(b) Spectrum $S(\omega)$ for $\xi =6.1$. This curve corresponds to a cut along the arrows in (a).
	(c) Same as in (a) but with $\Omega_j$ and $\mathcal{J}_j$ following a Gaussian
    distribution with a standard deviation of $2\%$ around their mean values. 
	(d) Cut along the arrows in (c).
	}
	\label{fig:4}
\end{figure} 
For a nonuniform TFIC, no closed analytical expressions for $\tilde{\rho}
(\omega)$ are available. Thus, we have to consider finite system sizes and calculate
numerically the relevant quantities, specifically, the spectrum of a finite-size
nonuniform TFIC,
\ma{
\tilde{\rho} ( \omega) = 2 \pi \sum_k \phi_{k,1}^2 \delta(\omega - \Lambda_k),
\label{eq:spectrum_finite_size}
}
which is the Fourier transform of \Eref{eq:autocorrelator}. To take the effect of
qubit decay processes into account, we phenomenologically broaden the delta-peaks in
\eref{eq:spectrum_finite_size} and replace them by Lorentzians of width $\gamma$
around the $\Lambda_k$. We model the nonuniformity of the qubit
parameters by writing $\Omega_j = \Omega \tau_j$ and $ \mathcal{J}_j = \mathcal{J}
\tau^\prime_j$ and choosing $\tau_j$ and $\tau_j^\prime$ to be random variables, which
follow Gaussian distributions with means $1$ and standard deviations $\sigma_\tau =
\sigma_{\tau^\prime} = 0.02$. 
Uniformity of the qubit parameters $\Omega_j$ and
$\mathcal{J}_j$ of this degree will turn out
to be sufficient for all proposed experiments. Much stronger disorder is not generally 
tolerable, as we will see below. However, from the experimental data for a
sample with three (even spatially separated) qubits presented in \cite{Fink2009}, 
we calculate a standard deviation of the qubit transition frequencies 
from their mean of $0.8\%$ (for zero flux bias). Thus, the requirements 
on the uniformity of $\Omega_j$ and $\mathcal{J}_j$ appear to be attainable.
With a typical set of system parameters that was also
used in \cite{Viehmann2012}, we numerically calculate $\tilde{\rho}(\omega)$ according to
\eref{eq:spectrum_finite_size} and the corresponding resonator spectrum $S$ 
according to \eref{eq:Scoupled}. In order to judge the effects of disorder, we also
reproduce our calculation of  $S$ for the corresponding uniform system
\cite{Viehmann2012} (Fig.\ S6). \Fref{fig:4} shows $S$ as a
function of $\omega$ and the (mean) normalized transverse field $\xi= \Omega/
2\mathcal{J}$ for the uniform system [figures \ref{fig:4}(a,b)] and for a typical disorder
configuration [figures \ref{fig:4}(c,d)]. In the uniform case, the signatures of
the QPT at $\xi=1$, the dispersive shift of the resonator frequency, and, on resonance, the double-peak 
with a separation of $4J$ (rather than $2g$ as in the
case $N=1$) that we have discussed in detail for $N\rightarrow \infty$ in \cite{Viehmann2012}
are clearly visible also for $N=20$. These characteristic features are insensitive
with respect to a small amount of disorder in the system parameters, as figures
\ref{fig:4}(c,d) demonstrate. 

We remark that in several recent circuit QED experiments the qubits have been found to be
unexpectedly hot \cite{Corcoles2011,Murch2012,Geerlings2012}. 
A corresponding non-negligible equilibrium population of the excited many-body eigenstates of the Ising chain
in our setup would lead to additional lines in the described spectroscopy experiment, at 
frequencies smaller than the bandwidth of the Ising chain. In the experimentally realistic case
that the Ising chain is deeply in the paramagnetic phase ($\Omega \gg 2J$), these resonances
at $\omega \lesssim 4J$ (the bandwidth of the chain in the paramagnetic phase) would be
well below the lower band edge $\Omega - 2J$. Thus, they would be distinguishable from the band of the Ising chain as
plotted in \fref{fig:4}, and their intensity might allow one to estimate the spurious
population of the excited states. However, for the proposed time-domain experiments with our circuit QED quantum
simulator that we discuss in the following sections, a non-negligible equilibrium
excitation of the Ising chain might necessitate post selection or initialization techniques.

\section{Disorder effects on the system dynamics}
A particularly interesting application of the proposed system would be to simulate
the nonequilibrium dynamics of the TFIC. In \cite{Viehmann2012}, we have suggested to
experimentally track the propagation of a localized excitation in the (uniform) TFIC that can
be easily created in our system and to measure the system dynamics after quenching
the transition frequencies of all qubits. In this section, we show that none of the predicted
experimental results changes qualitatively if the parameters of the TFIC are
slightly disordered, as has to be expected in reality. Stronger disorder,
accessible, e.g., by deliberately detuning individual qubits, is shown to produce
qualitatively different physics in the previously proposed experiments,
like Anderson localization of the propagating excitation. 
For the realistic case $\Omega \gg J$, we give an estimate of the corresponding
disorder strength. Finally, we develop a quantitative
theory of the effects of weak disorder on the system's nonequilibrium dynamics that 
explains the results of numerical experiments with the disordered TFIC. This theory might be
helpful for experimentalists to estimate system and disorder parameters for
successfully performing nonequilibrium experiments with the TFIC (e.g., for a given
measurement resolution) without having
to do numerical simulations.

\subsection{Propagation of localized excitations}\label{subsec:loc_excitation}
\begin{figure}
	\centering
	\includegraphics[width=\columnwidth]{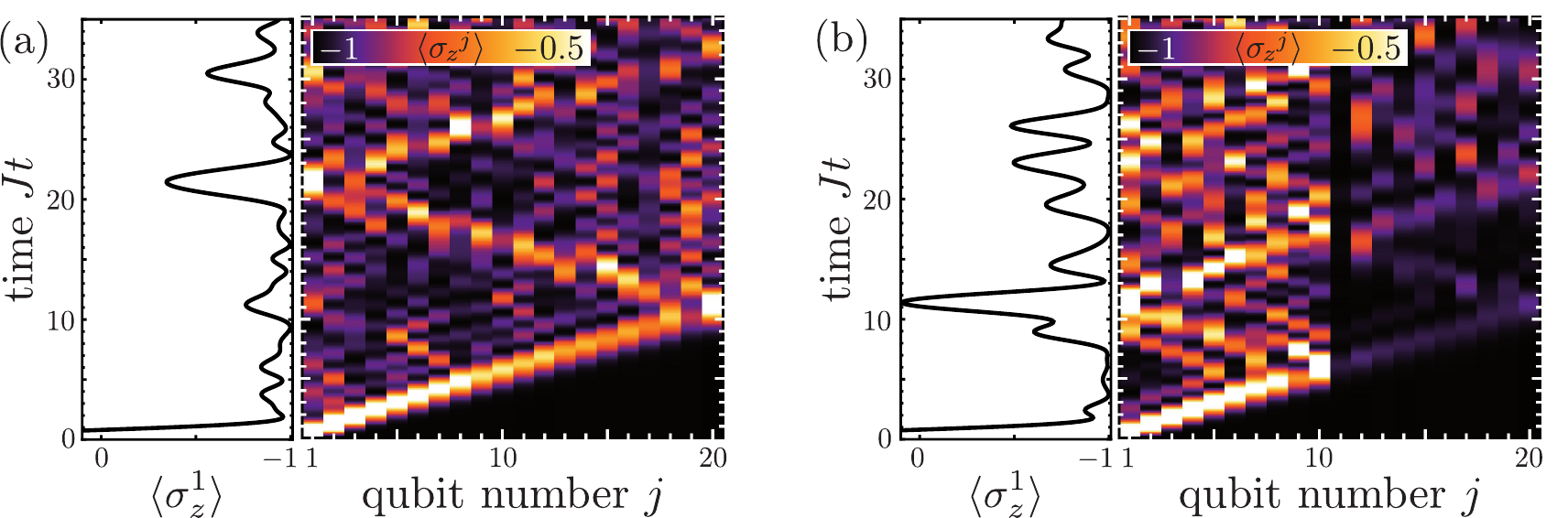}
	\caption{
    Propagation of a localized excitation in a slightly disordered transverse-field
    Ising chain of length $N=20$. Specifically, the density plots show the nonequilibrium
    time evolution of $\langle \sigma_z^j \rangle $ for all $j$ after a
    $\pi$-pulse on the first qubit while the system is in the paramagnetic phase (mean
    normalized transverse field $\xi = 8$). For better visibility of the features,
    values $>-0.5$ are plotted white. The experimentally
    accessible observable $\langle \sigma_z^1 \rangle$ is singled out in the left
    panels.
	(a) The qubit transition frequencies $\Omega_j$ and qubit-qubit couplings $\mathcal{J}_j$ are randomly chosen
	according to Gaussian distributions with standard deviations of $2\%$ from the
    mean values.
	(b) Same as in (a) but with qubit $11$ strongly detuned.}
	\label{fig:5}
\end{figure}

For the first type of experiments we have suggested in \cite{Viehmann2012}, it is assumed
that the TFIC is deeply in the paramagnetic phase ($\xi \gg 1$) and detuned from the resonator.
In this situation, the TFIC is essentially decoupled from the resonator and its ground
state is characterized by $\langle \sigma_z^j\rangle \approx -1$. Applying a fast $\pi$-pulse to the first qubit thus
creates a localized excitation in the system that subsequently propagates trough the chain due to
the qubit-qubit coupling $\mathcal{J}$. The time evolution of the observable $\langle \sigma_z^j
\rangle$  after the $\pi$-pulse can be approximately described by \cite{Viehmann2012}
\begin{eqnarray}
\langle \sigma_z^j \rangle (t) =
& - \sum_k \psi_{k,j} \phi_{k,j} + \sum_{k, k^\prime} \rme^{i (\Lambda_k - \Lambda_{k^\prime})t}  \big[ \phi_{k,1} \phi_{k^\prime,1} \nonumber \\
&\times  (  \psi_{k,j} \phi_{k^\prime,j} + \psi_{k^\prime,j} \phi_{k,j}) \big]. \label{eq:spin_wave_explicit}
\end{eqnarray}
We plot this result in \fref{fig:5}(a) for all $j$ in a chain of length $N=20$ and
for a mean normalized transverse
field $\xi = \Omega/ 2\mathcal{J} = 8$ (same system parameters as in
\cite{Viehmann2012}), and, again, we randomly choose $\Omega_j = \Omega \tau_j$ and
$\mathcal{J}_j= \mathcal{J} \tau_j^\prime$ according to
Gaussian distributions with standard deviations of $2\%$ from the mean values $\Omega$
and $\mathcal{J}$ as before (right panel). The experimentally measurable observable $\langle
\sigma_z^1 \rangle (t)$ is singled out in the left panel. The propagation of a
localized excitation through the chain, and its reflection at the far end of the chain
that leads to a distinct revival of $\langle \sigma_z^1 \rangle(t)$ at $t\approx
N/J$, are still clearly visible in this slightly nonuniform system. 

If the transition frequencies $\Omega_j$ of the qubits can be tuned individually, the
effective length of the TFIC has been shown to be adjustable by strongly detuning one qubit from the
others \cite{Viehmann2012}. This holds true also for a slightly nonuniform system:
\Fref{fig:5}(b) shows a typical result for a system with the same parameters and disorder strength as in (a),
but with qubit $11$ strongly detuned by setting $\tau_{11}=1.3$. This result is
qualitatively identical with the result for the corresponding non-disordered system
\cite{Viehmann2012}. The strong nonuniformity at $j=11$ acts as a barrier for the
propagating excitation and leads to its reflection. Thus, it effectively changes
the length of the TFIC. 

Having shown that the experiments with propagating localized excitations proposed in
\cite{Viehmann2012} yield qualitatively the same results for ordered and slightly
disordered systems, we now proceed and study disorder effects on this type of
experiments quantitatively. Parts of the following analysis also
apply to other nonequilibrium experiments with the TFIC, as will be discussed in the
context of quantum quenches (\sref{subsec:quantum_quenches}).

\begin{figure}
	\centering
	\includegraphics[width=\columnwidth]{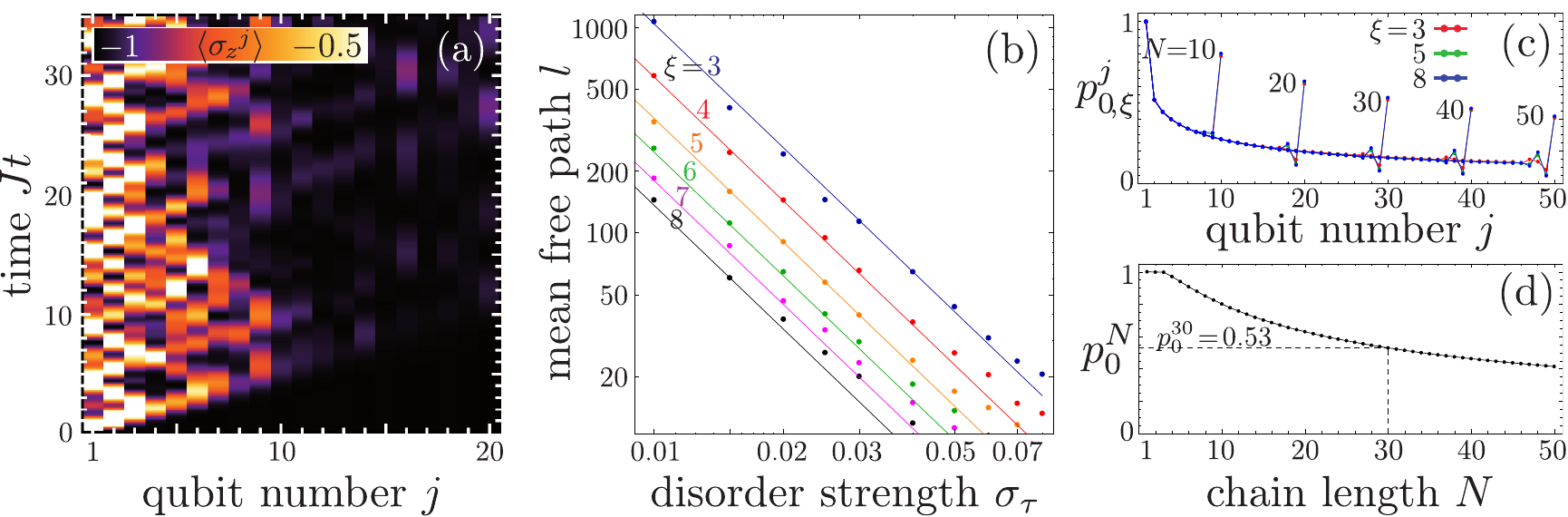}
	\caption{
    (a) Propagation of an initially localized excitation in a strongly
    disordered transverse-field Ising chain. Initialization and system parameters are
    identical to
    \fref{fig:5}(a), but the $\Omega_j$ and $\mathcal{J}_j$ are randomly chosen
    according to Gaussian distributions with standard deviations of $6.5\%$ from the
    mean values. The plot clearly shows that strong localization of the excitation
    prohibits its propagation through the chain.
    (b) Mean free path $l$ of the propagating excitation (defined in the main text)
    vs. normalized standard
    deviation $\sigma_\tau$ of the qubit transition frequencies for different values
    of the normalized transverse field $\xi$ on log-log scale.
    The points are $l_{\sigma_\tau, \xi}$ as gained by numerically 
    averaging many disorder configurations. The lines are best fits of
    $1/\sigma_\tau^a \xi^b$ to these data. 
    (c,d) Behavior of a non-disordered system, with uniform $\Omega_j =\Omega$ and
    $\mathcal{J}_j= \mathcal{J}$, for comparison.
    (c) Maximum excitation probabilities $p_{0,
    \xi}^j$ of the $j$th qubits in the nonequilibrium time evolution of 
    uniform transverse-field Ising chains of lengths
    $N=10,20,30,40,50$ after the first qubit has been flipped. For each chain length,
    $p_{0,\xi}^j$ is plotted for $\xi = 3,5,8$ (red, green, blue). Apart from
    boundary effects, the decay of
    $p_{0,\xi}^j$ with $j$ is slower than $\propto 1/j$. The maximum excitation
    probabilities $p_{0,\xi}^N$ of the last qubits of the chains are significantly
    enhanced compared to nearby bulk sites.
    (d) Maximum excitation probability $p_{0}^N$ of the $N$th qubit vs. chain length $N$
    (for any $\xi \gg 1$).}
	\label{fig:6}
\end{figure}

Since it is assumed that the system is deeply in the paramagnetic phase, the
mean qubit transition frequency $\Omega$ is larger than the modulus of the mean qubit-qubit coupling
$J$, $\Omega/ J \gg 1$. As before, we further assume uncorrelated
disorder of the system parameters via $\Omega_j= \Omega \tau_j  $ and
$\mathcal{J}_j = \mathcal{J} \tau^\prime_j $, where $\tau_j$ and
$\tau^\prime_j$ follow Gaussian distributions with standard 
deviations $\sigma_\tau$ and $\sigma_{\tau^\prime}$ from $1$. That is, for
$\sigma_\tau = \sigma_{\tau^\prime}$, the absolute variation of the $\Omega_j$ will
be larger than the absolute variation of the $\mathcal{J}_j$. Therefore, the dynamics of the
system may be expected to be much more sensitive to increasing $\sigma_\tau$ than
$\sigma_{\tau^\prime}$. Moreover, one may expect that disorder effects start to
qualitatively affect the system dynamics even of small systems (that is, on the scale
of neighbouring qubits $j$ and $j+1$) when the disorder in the qubit transition
frequencies becomes comparable to the modulus of the mean qubit-qubit coupling, 
$\Omega \sigma_\tau = J$. These deliberations are confirmed by numerical experiments: We
first consider the wave functions $g_{k,j}$ and $h_{k,j}$ in position space ($\eta_k
= \sum_{j=1}^N g_{k,j} c_j + h_{k,j} c_j^\dagger$). For zero disorder, they are
extended over the whole chain (except for the mode with $\Lambda_k \rightarrow 0 $ in
the ordered phase \cite{Kitaev2001}). Increasing $\sigma_\tau$ localizes the wave
functions much more strongly than increasing $\sigma_{\tau^\prime}$, and the localization length of the
wave functions indeed reduces from many ($\gg 1$) sites to a few ($\gtrsim 1$) sites
at $\Omega \sigma_\tau \approx J$. 
Correspondingly, the propagation of an excitation
initially localized at site $1$ is only weakly affected by disorder in $\mathcal{J}$.
However, if $\sigma_\tau \gtrsim J/\Omega$, it propagates only a few sites before becoming completely
trapped due to the disorder. This manifestation of Anderson
localization \cite{Anderson1958} is illustrated in \fref{fig:6}(a), where we have
used the same system parameters as in \fref{fig:5}, but we have randomly chosen
$\tau_j$ and $\tau^\prime_j$ according to Gaussian distributions around $1$ with
standard deviations $\sigma_\tau = J/\Omega = 0.0625= \sigma_{\tau^\prime} $. For
definiteness, we always choose $\sigma_{\tau^\prime} = \sigma_\tau$ in the following.

We have seen that for $|\xi |\gg 1$ (paramagnetic phase) the effective disorder
strength in the quantum Ising chain is set by $\sigma_\tau \Omega/ J\propto
\sigma_\tau| \xi|$. Now we try to determine how the relevant observables
in the currently considered type of experiments depend on this quantity.
The observable we focus on in the following is the maximum excitation probability
(maximized over time) of
the $j$th qubit caused by the propagation of the localized excitation through the disordered
chain. In an experiment, one would for instance create an excitation of the first qubit and measure
the excitation probability of some other (e.g.\ the $N$th) qubit as a function of time. The maximum
excitation probability of the $j$th qubit is an important quantity since it will
determine if the effect of the propagating excitation can be measured at site $j$,
given a certain measurement resolution. In a single disordered system, the maximum excitation
probability of qubit $j$ will depend on the specific (random) disorder configuration
of this system. Therefore, a study of the effect of disorder as characterized by
the statistical quantity $\sigma_\tau |\xi|$ can only refer to the statistical average of the
maximum excitation probability of qubit $j$ in one disordered system 
over an ensemble of many disordered systems (disorder configurations),
all chosen according to the same probability distribution. Stated as a formula, this ensemble
average of the maximum excitation probability of qubit $j$ is given by
\ma{
p^j_{\sigma_\tau,|\xi|} = \frac{1}{2} \Big(\,\overline{\overline{\max_t [ \langle \sigma_z^j
\rangle(t)]}} +1\Big).
}
Here, the double overbar $\overline{\overline{\,\cdot\,}}$ denotes the ensemble
average over many disordered systems (disorder configurations) with the same system and
disorder parameters $\xi$, $J$, and $\sigma_\tau = \sigma_{\tau^\prime}$. This
average is taken after one has maximized $\langle \sigma_z^j \rangle (t)$ for a specific
disordered system over time. Our goal is to find the explicit functional dependence of 
$p^j_{\sigma_\tau,|\xi|}$ on $\sigma_\tau$ and $|\xi|$ (in fact, we expect dependence only on the
product $\sigma_\tau |\xi|$). Note that we assume that $p^j_{\sigma_\tau,|\xi|}$ depends neither on
the sign of $\xi$ nor explicitly on the mean qubit-qubit coupling $J$, but only on the ratio of
$\Omega$ and $J$ (via $|\xi|$). This is strictly true for $\sigma_\tau
=\sigma_{\tau^\prime} = 0$. By explicitly solving equations \eref{eq:relpsiphi} for
this case \cite{Viehmann2012}, one can show that after substituting $\xi \rightarrow -\xi$,
the new allowed wave vectors are
$q=\pi-k$ with $\Lambda_q = \Lambda_k$, $\phi_{q,j} = (-1)^{N-j} \phi_{k,j}$, and
$\psi_{q,j} = (-1)^{N-j} \psi_{k,j}$. With that one can easily see that
equation \eref{eq:spin_wave_explicit} does not depend on the sign of $\xi$. Moreover, 
$\phi_k$ and $\psi_k$ are independent of $J$ [which also follows from equations
\eref{eq:relpsiphi}], and $\Lambda_k \propto J$ such that changing $J$
corresponds only to a rescaling of time. The influence of disorder, however, is
essentially set by $\sigma_\tau |\xi|$ (for $|\xi| \gg 1$), as we have argued above.
Consequently, we may take $p^j$ to be independent of $J$ and of the sign of $\xi$.
Nevertheless, to keep notation short, we write $\xi$ instead of $|\xi|$ for the
remainder of this chapter. For simplicity, we first focus
on a semi-infinite system ($N \rightarrow \infty$) and discuss later the 
increase of $p^j$ at the end of the chain (due to the refocusing of the
dispersed wave packet of the propagating excitation).

As usual for disordered systems (e.g.\ \cite{Akkermans2007}), we will try to characterize 
the disorder effects on the ensemble-averaged maximum qubit excitation 
$p^j_{\sigma_\tau,\xi}$ via a mean free path. To that end, it
pays to first discuss in more detail the uniform case, $p^j_{0,\xi}$.
Even there, analyzing the propagation of the dispersive wave packet that determines
the maximum excitation probability of a qubit requires some care.
For $\xi \gg 1$, this excitation probability does not depend on
$\xi$. This is because the dispersion relation of the TFIC becomes that of the tight binding
model, $\Lambda_k= 2 J \sqrt{1 + \xi^2 - 2 \xi\cos k }\approx 2J \mathrm{sign}(\xi)
(\xi - \cos k)$. Thus, $\xi$ only sets the band gap but
does not influence the shape of the dispersion relation.
Except for the aforementioned boundary effects,
$p^j_{0,\xi}$ also does not depend on $N$. This is evident from \Fref{fig:6}(c), where we plot
$p^j_{0,\xi}$ for several $\xi$ and $N$. The curves for different $\xi$ but same $N$
lie almost on top of each other (henceforth, we drop the index $\xi$ form
$p^j_{0,\xi}$), and curves for different $N$ can be distinguished
only by the boundary effects, that is, by the strong increase of $p_{0}^j$ at
$j= N$ (which will be discussed later).
The decay of the $p^j_{0}$ with $j$ is relatively slow (slower than $1/j$),
which should considerably simplify the experiments proposed in \cite{Viehmann2012}.
This slow decay of $p^j_{0}$ can be understood from the dispersion relation
$\Lambda_k$ of the system which, for $\xi \gg 1$, is quadratic in $k$ at $k \approx
0, \pi$, and linear at $k \approx \pi/2$: If an initially localized 
wave packet with width $s$ and momentum $q$, $\psi (x,0)= \alpha \rme^{-x^2/2 s^2 + i q x}$,
$\alpha = (s^2 \pi)^{-1/4}$, is evolved in time by the Hamiltonians
$H_1 = h_1 k$ and $H_2 = h_2 k^2$, respectively, one finds 
\begin{eqnarray}
|\psi(x,t)|^2_{H_1}& = |\psi(x-h_1 t,0)|^2 = \alpha^2 \rme^{-(x-h_1 t)^2/s^2},\\
|\psi(x,t)|^2_{H_2}& = \frac{ \alpha^2 s^2 }{\sqrt{s^4 + 4 h_2^2 t^2}} \exp \left(-
\frac{(x-2 h_2 t q)^2}{s^4 +4h_2^2 t^2} \right).
\end{eqnarray}
That is, for $H_1$, maximum and width of the probability distribution for
finding the particle at a position $x$ are constant, while for $H_2$ and strong
initial localization (or large times) the width is $\propto t$ and the maximum
$\propto 1/t$. As the dispersion relation of the TFIC interpolates between these two
cases, one may expect a decay of $p^j_{0}$ slower than $1/j$. 

Coming back now to the disordered case, one might suspect that the ensemble-averaged
maximum qubit excitation $p^j_{\sigma_\tau, \xi}$ is related to the corresponding
quantity for a non-disordered system $p^j_0$ via an exponential decay, governed by a finite mean free path 
$l_{\sigma_\tau , \xi}$ for the propagation of the localized excitation,
\ma{
p^j_{\sigma_\tau, \xi} = p^j_{0} \rme^{-j/l_{\sigma_\tau ,
\xi}}.\label{eq:dep_MFP}
}
If \eref{eq:dep_MFP} holds, 
\ma{
\frac{1}{l_{\sigma_\tau , \xi}} = \frac{1}{j} \ln \left(
\frac{p^j_{0}}{p^j_{\sigma_\tau, \xi}} \right) \label{eq:inverse_MFP}
}
should be independent of $j$. This observation can be used to check our assumption
\eref{eq:dep_MFP}. We numerically calculate $p^j_{\sigma_\tau, \xi}$ for
all combinations of $\xi = 3, \ldots, 8$ and $100 \times \sigma_\tau \in
\{1,1.5,2,2.5,3,4,5,6,7,8\}$ in a chain of length $N=20$, and we average over $100$ disorder
configurations. This turns out to be a good compromise between calculation time and
ensemble and system size as long as the effective disorder $\sigma_\tau \xi$ is not too small (see
below). With these $p^j_{\sigma_\tau, \xi}$, we calculated the RHS of
\eref{eq:inverse_MFP} for $j=5,\ldots,16$. Other $j$ are not considered, in order to
minimize boundary effects. Our results for $j=5,\ldots,16$ are approximately equal,
with the ratio of standard
deviation to mean value being $<0.1$ for given $\sigma_\tau$ and $\xi$. We note that
for very weak effective disorder $\sigma_\tau \xi \lesssim 0.1$ we have to average
over 500 disorder configurations such that this ratio is $<0.1$, because with
decreasing ratio
$ p^j_{0}/ p^j_{\sigma_\tau, \xi}$ the slope of the logarithm on the
RHS of \eref{eq:inverse_MFP} increases.  
Thus, the numerical data seem to confirm our assumption \eref{eq:dep_MFP}, and 
the influence of disorder on the considered experiment is captured by a
mean free path $l_{\sigma_\tau , \xi}$. In our subsequent analysis, we try to find simple
expressions for this quantity.
 
The propagation of the localized excitation in the Gaussian disordered TFIC is akin to the propagation of
a particle in an uncorrelated random potential $V(r)$ with $\langle V(r) V(r^\prime)
\rangle = V_0^2 \delta(r-r^\prime)$. To lowest order in perturbation theory
(Fermi's golden rule, e.g.\ \cite{Akkermans2007}), the mean free path of the latter
decreases as the inverse square of the disorder strength, $\propto 1/V_0^2$. In our
case, the effective disorder strength is determined by the dimensionless quantity $\sigma_\tau \xi$. 
Therefore, we expect that
\ma{
l_{\sigma_\tau, \xi} = \frac{1}{(\sigma_\tau \xi)^2}. \label{eq:MFP_closed_form}
}
To check this, we calculate $l_{\sigma_\tau , \xi}$ for the same combinations of $\sigma_\tau$ and
$\xi$ as before by averaging the RHS of \eref{eq:inverse_MFP} over $j=5,\ldots, 16$.
Then we fit the function $l(\sigma_\tau , \xi) =1/ \sigma_\tau^{a} \xi^{b} $
to our data for $l_{\sigma_\tau , \xi}$. We find the exponents $a \approx 2.002$ and $b\approx
2.071$, which comes close to our expectation of $a=b=2$. Numerical data and fit are plotted on
log-log scale in \fref{fig:6}(b). As long as
the effective disorder strength is not too big ($\sigma_\tau \xi \lesssim 0.2$),
$l(\sigma_\tau , \xi)$ with the fit values of $a$ and $b$ reproduces the numerically
(by ensemble-averaging) extracted mean free path $l_{\sigma_\tau , \xi}$. 
Here, one may attribute the deviations of $a$
and $b$ from $2$ to the finite ensemble sizes. For stronger disorder, however, the
fit of $l(\sigma_\tau, \xi)= 1/\sigma_\tau^{a} \xi^{b}$ begins to deviate from
$l_{\sigma_\tau , \xi}$. Thus, higher-order effects (beyond Fermi's golden rule)
and/or the disorder in $J$ seem to be no longer negligible. 

Finally, in setups with a second readout resonator (cf. \fref{fig:1}),
the maximum population $p_{\sigma_\tau,\xi}^N$ of the $N$th
qubit will be an experimentally relevant quantity. Since the dispersed wave packet of
the propagating excitation is refocussed at the end of the chain, the maximum
excitation probability of the $N$th qubit is considerably enhanced compared to nearby
bulk qubits [see \fref{fig:6}(c)]. It turns out
that $p_{\sigma_\tau,\xi}^N$ can also be estimated by means of \eref{eq:dep_MFP},
the mean free path \eref{eq:MFP_closed_form},
and the value of $p^N_{0}$ for the corresponding non-disordered system, which we
plot for $N=1,\ldots,50$ in \Fref{fig:6}(d).

\begin{figure}
	\centering
	\includegraphics[width=\columnwidth]{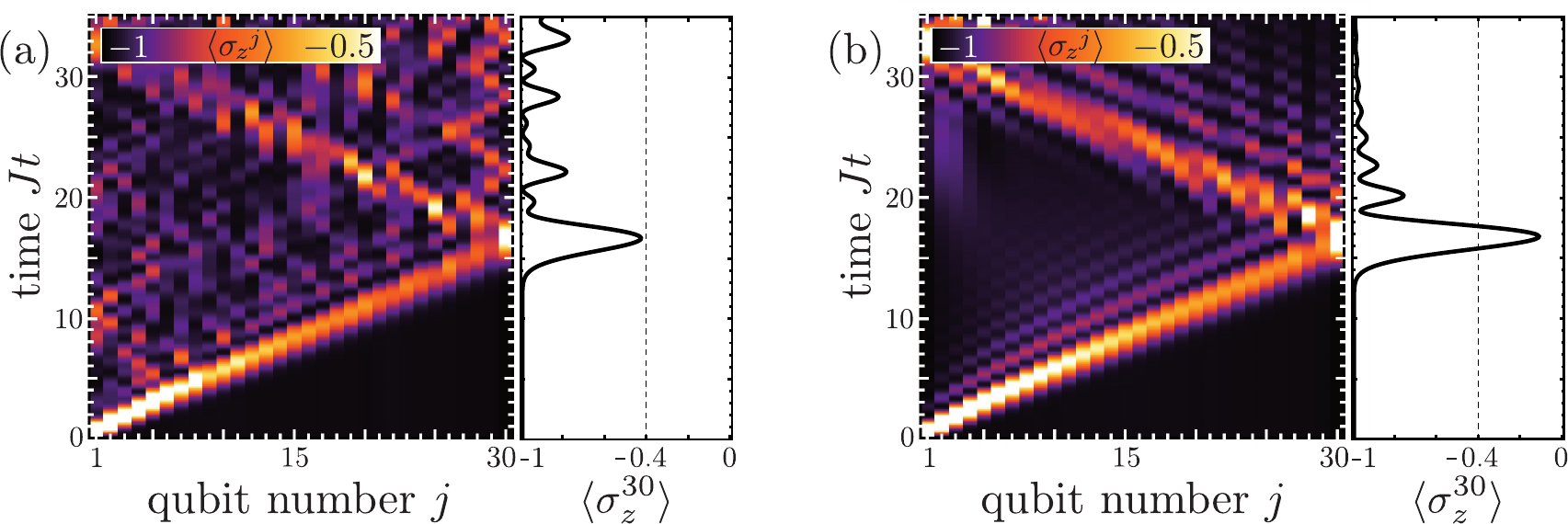}
	\caption{
    Propagation of a localized excitation in a nonuniform transverse-field
    Ising chain of length $N=30$ and with normalized transverse field $\xi=4$. 
	(a) The qubit transition frequencies $\Omega_j$ and qubit-qubit couplings $\mathcal{J}_j$ are randomly chosen
	according to Gaussian distributions with standard deviations of $3.4\%$ from the
    mean values.
	(b) Same as in (a) but with a cosine modulation of the qubit transition
    frequencies $\Omega_j$ with standard deviation $\approx 3.4 \%$, instead of
    uncorrelated disorder of $\Omega_j$ and $\mathcal{J}_j$.}
	\label{fig:7}
\end{figure} 
Summing up, equations \eref{eq:dep_MFP} and \eref{eq:MFP_closed_form},
together with Figures \ref{fig:6}(c,d) allow one to easily
estimate suitable system and disorder parameters for
successfully implementing the presently considered type of experiment. For instance, if in a system with
$N=30$ and $\xi = 4$ the $N$th qubit should get a population of $p_{\sigma_\tau,
\xi}^{30} = 0.3$ (which corresponds to $\max \left[ \langle \sigma_z^{30} \rangle
(t)\right] = -0.4$),
then one can roughly (i.e., averaged over many systems) afford a standard deviation of the qubit
transition frequencies from their mean of $\sigma_\tau = \left[(N \xi^2)^{-1} \log
\left( p_0^{30}/0.3 \right) \right]^{1/2} \approx 0.034$,
where we have extracted $p_0^{30} \approx 0.53$ from \fref{fig:6}(d). 
A typical result for these parameters is plotted in \fref{fig:7}(a). Here, the maximum
excitation probability is found to be $p_{0.034, 4}^{30} \approx 0.29$ (since $\max_t \left[
\langle \sigma_z^{30} \rangle (t)\right] \approx -0.43$).

We remark that the foregoing deliberations only hold for uncorrelated disorder
of the system parameters and do not take into account qubit decay. Correlated
disorder can yield qualitatively different results and has to be studied explicitly
via equation \eref{eq:spin_wave_explicit}. We also remark that if the $\Omega_j$ are
individually tunable, it becomes possible to study the propagation
of localized excitations in arbitrary potentials. For instance, it might be interesting to
choose $\Omega_j = \Omega [1 +\sqrt{2}\sigma_\tau \cos( 2\pi j/N)]$ and to compare
the system dynamics with the Gaussian disordered case. For large $N$, both
distributions of $\Omega_j$ have the same mean and the same standard deviation, but
in the former case the system is not disordered and the localization of the
propagating excitation is much weaker than in the genuinely disordered case.
\Fref{fig:7}(b) shows the propagating excitation in such a system with $N$, $\xi$, and
$\sigma_\tau$ as in \fref{fig:7}(a) (with uniform $\mathcal{J}_j$).

\subsection{Quench dynamics} \label{subsec:quantum_quenches}
The second type of nonequilibrium experiments we have proposed in \cite{Viehmann2012}
relies on the possibility to rapidly change the transition frequency
$\Omega$ of a superconducting qubit in a circuit QED system by tuning the magnetic
flux through its SQUID loop. This has been shown to be possible
virtually instantaneously on the dynamical time scale of a circuit QED system 
\cite{DiCarlo2010,Fedorov2012,Reed2012}, without changing the system's wave function.
Let us now assume that the circuit QED quantum simulator of the (uniform)
TFIC proposed in \cite{Viehmann2012} is implemented with Cooper-pair boxes. For this system, such a sudden sudden change of all
$\Omega_j = \Omega$ corresponds to a global quantum quench of the normalized transverse magnetic field $\xi =
\Omega/2 \mathcal{J}$. We remark that one can also produce quenches of 
$\xi$ by using transmons in a non-standard parameter regime
instead of Cooper-pair boxes, or by using usual transmons with tunable coupling capacitances
\cite{Averin2004,Viehmann2012}. 
We also remark that the observation of the phenomena described in the following will
set higher requirements
on the energy relaxation and phase coherence times of the collective many-body quantum states of the Ising chain than the experiments
proposed in \sref{subsec:Disorder_Spectrum} and \sref{subsec:loc_excitation}. The global
quantum quench brings the Ising chain in a globally excited state whose time evolution has to
be coherent on the time scale $ N/J$ of these phenomena (see below). 
Nevertheless, meeting this constraint seems feasible, since even for $N=30$ and a moderate
coupling strength $J/2\pi = 100$ MHz, we find $N/J \sim 50 $ns, which is far below the energy
relaxation times $T_1 \sim 7.3 \mu$s and coherence
times $T_2 \sim 500$ns achieved for individual Cooper-pair boxes \cite{Wallraff2005}.

The nonequilibrium dynamics of the TFIC following
a quantum quench is currently subject to much theoretical research, e.g.\
\cite{Polkovnikov2011,Barouch1970,Igloi2000,Calabrese2006,Rossini2009,Igloi2011,Calabrese2011,Rieger2011,Heyl2012,Calabrese2012a,Calabrese2012b,Marino2012,Essler2012}, and should be
experimentally observable with our circuit QED quantum simulator. 
In this context it is usually assumed that for $t<0$ the system is in the ground state $\ket{0}_a$ of
a Hamiltonian $\mathcal{H}_{\mathrm{I},a}$ (characterized by $\xi_a$).
At $t=0$, the overall transverse field is changed, $\xi_a
\rightarrow \xi_b$, and the nonequilibrium time evolution of some observable
$\mathcal{O}$ under $\mathcal{H}_{\mathrm{I},b}$ is investigated,
\ma{
\langle \mathcal{O} \rangle (t) = _a\!\!\bra{0}\rme^{\rmi t \mathcal{H}_{\mathrm{I},b}}
\mathcal{O} \rme^{-\rmi t  \mathcal{H}_{\mathrm{I},b} } \ket{0}_a .
\label{eq:quench}
}
In \cite{Viehmann2012} we have
focussed on the time evolution of the local transverse magnetization $\langle
\sigma_z^j \rangle$ and the end-to-end correlator $\langle \sigma_x^1 \sigma_x^N
\rangle$ (indicating long-range order) after quenching $\xi$ within the paramagnetic
phase. These quantities should be experimentally easily accessible in our system. 
In this section, we show that also for such quantum quenches the predicted
experimental results of our earlier work are insensitive to
a small amount of fabrication-induced disorder. 

In general, two sets of the $\Omega$- and $\mathcal{J}$-parameters, $\{\Omega^{a/b}_j \}$
and $\{ \mathcal{J}^{a/b}_j \}$, fully specify the Hamiltonians
$\mathcal{H}_{\mathrm{I},a/b}$ [equation \eref{eq:TFIC}]. Given these parameters, the
time evolution \eref{eq:quench} of the local magnetization and the end-to-end correlator
can be written as \cite{Viehmann2012}
\begin{eqnarray}
\langle \sigma_z^j \rangle (t) = -\sum_k& \psi_{k,j}^b \phi_{k,j}^b + 2
\sum_{k,k^\prime} \lbrace \psi_{k,j}^b \phi_{k^\prime,j}^b \times \nonumber \\
&[
X_{k,k\prime} \cos t(\Lambda_k^b + \Lambda_{k^\prime}^b) + 
Y_{k,k\prime} \cos t(\Lambda_k^b - \Lambda_{k^\prime}^b)
] \big\rbrace ,\label{eq:loc_mag_quench}\\
\langle \sigma_x^1 \sigma_x^N \rangle (t) = \sum_k & \phi_{k,1}^b \psi_{k,N}^b  + 2
\sum_{k,k^\prime} \lbrace \phi_{k,1}^b \psi_{k^\prime,N}^b  \times \nonumber \\
&[
X_{k,k\prime} \cos t(\Lambda_k^b + \Lambda_{k^\prime}^b) - 
Y_{k,k\prime} \cos t(\Lambda_k^b - \Lambda_{k^\prime}^b)
] \big\rbrace .\label{eq:endtoend_quench}
\end{eqnarray}
Here,   
\begin{eqnarray}
X_{k,k\prime}&=\big[(g_k^b)^T H^a + (h_k^b)^T G^a \big] \big[ (G^a)^T
g_{k^\prime}^b + (H^a)^T h_{k^\prime}^b\big], \\
Y_{k,k\prime}&=\big[ (g_k^b)^T H^a + (h_k^b)^T G^a \big] \big[ (H^a)^T
g_{k^\prime}^b + (G^a)^T h_{k^\prime}^b \big], 
\end{eqnarray}
and $G$ and $H$ are matrices that respectively contain the $g_k$ and $h_k$ as columns.
In these equations, a quantity carrying the index $a$ or $b$ is to be calculated from
equations \eref{eq:relpsiphi} with parameter set $a$ or $b$.

To implement disorder of the system parameters before the quantum quench, we write again
$\Omega_j^a = \Omega^a \tau_j$ and $\mathcal{J}_j = \mathcal{J} \tau^\prime_j$, and
we randomly choose $\tau_j$ and $\tau_j^\prime$ according to Gaussian distributions
with standard deviations $\sigma_\tau$ and $\sigma_{\tau^\prime}$ from 1.
As we have argued in \sref{subsec:Implementation}, 
tuning the flux $\Phi$ through the SQUID loops of the qubits only changes the 
mean qubit transition frequency $\Omega^a \rightarrow \Omega^b$ (and, thus, the mean
transverse field $\xi_a = \Omega^a/2 \mathcal{J} \rightarrow \xi_b = \Omega^b / 2\mathcal{J}$),
but leaves $\tau_j$, $\mathcal{J}$, and $\tau_j^\prime$ unaffected. Hence, by fixing
$\xi_{a/b}$ and $\sigma_{\tau/\tau^\prime}$, the system is fully specified before and
after the quench (as in \sref{subsec:loc_excitation}, the absolute values of $\Omega^{a/b}$ and $\mathcal{J}$ can be
absorbed in the time scale $Jt$ of the dynamics),
and we are ready to evaluate equations \eref{eq:loc_mag_quench} and \eref{eq:endtoend_quench}. 

\begin{figure}
	\centering
	\vspace{1mm}
	\includegraphics[width=\columnwidth]{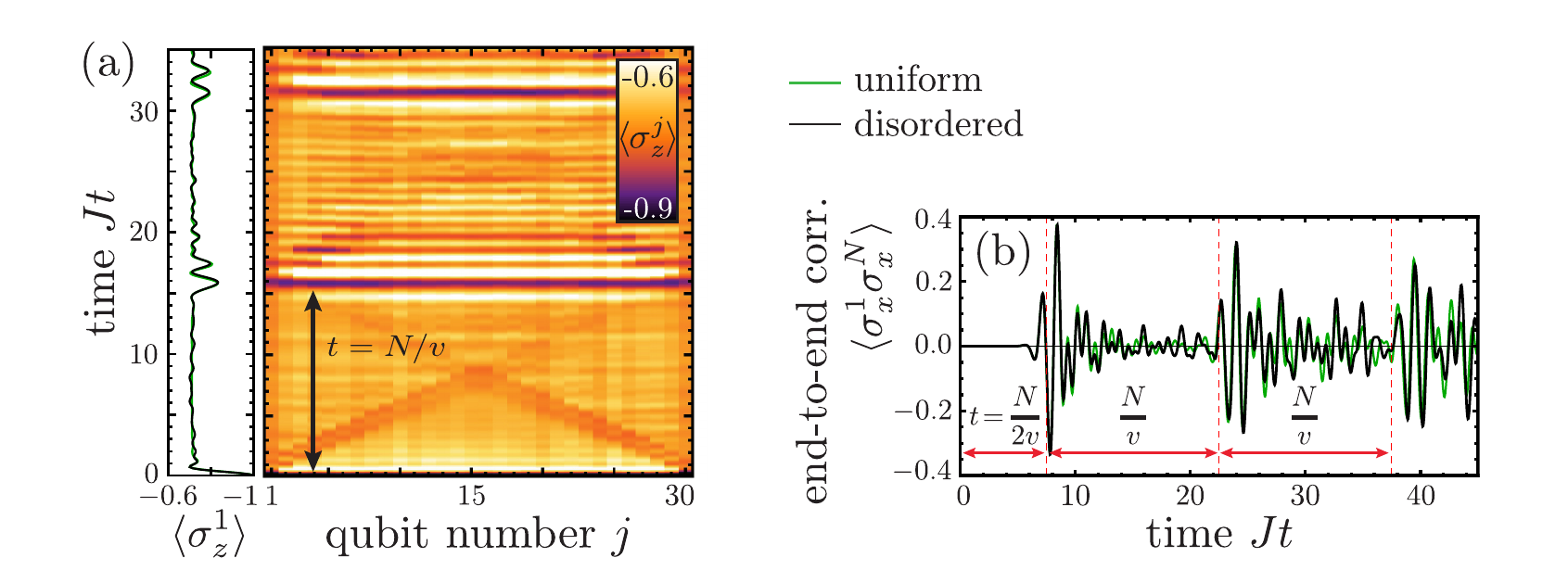}
	\caption{(a) Time evolution of the magnetization $\langle \sigma_z^j \rangle$ in a disordered TFIC of length $N=30$ 
	after a quench of the mean normalized transverse field $\xi = \Omega/ 2
    \mathcal{J} = 8 \rightarrow 1.2$
    (right). Values $<-0.9$ ($>-0.6$) are plotted black (white). The measurable observable $\langle \sigma_z^1 \rangle$ is plotted
    separately in the left panel (black), along with the corresponding trace for a
    uniform system (green). 
    (b) Time evolution of the end-to-end correlator $\langle \sigma_x^1 \sigma_x^N \rangle$ 
    in a disordered TFIC of length $N=30$ 
	after a quench of the mean normalized transverse field $\xi = 8 \rightarrow 1.5$
    (black), along with the corresponding trace for a uniform system (green). In both
    plots the qubit transition frequencies $\Omega_j$ and qubit-qubit couplings $\mathcal{J}_j$ are randomly chosen
	according to Gaussian distributions with standard deviations of $2\%$ from the
    mean values $\Omega$ and $\mathcal{J}$. 
    }
	\label{fig:8}
\end{figure}

\Fref{fig:8}(a) shows the local magnetization $ \langle \sigma_z^j \rangle (t)$ 
for all $j$ and for the same system parameters as in figure 4 of \cite{Viehmann2012},
but with $\Omega_j^{a/b}$ and $\mathcal{J}_j$ having standard deviations
$\sigma_\tau = \sigma_{\tau^\prime} = 2 \%$ around their mean values (right panel). The
experimentally easily measurable trace of $\langle\sigma_z^1\rangle$ is singled out in the left
panel (black). For comparison, we also plot (green) the local magnetization of the first
qubit $\langle \sigma_z^1 \rangle$ of the uniform system (as plotted in the left
panel of figure 4 of \cite{Viehmann2012}). Correspondingly, \fref{fig:8}(b) shows
\eref{eq:endtoend_quench} for a uniform system as in Figure S6 of \cite{Viehmann2012} (green), and  with 
$2\%$ disorder in $\Omega_j^{a/b}$ and $\mathcal{J}_j$ (black). The plots demonstrate that 
the quench dynamics of the considered observables is not qualitatively
affected by the presence of a small amount of disorder. 

For a more systematic analysis of the disorder effects on the quench experiments
considered here we make use of our findings for the mean free path of a propagating localized
excitation from the previous section. This is possible because the quench dynamics of the TFIC is
governed by the propagation of quasiparticles (QPs) through the system
\cite{Igloi2000,Calabrese2006,Igloi2011,Rieger2011,Sachdev1997,Viehmann2012}.
These correspond to flipped spins, essentially like the localized excitation of the previous
section. Indeed, if the system is initially in the paramagnetic phase, the time evolution immediately
after the quantum quench $\rme^{-it \mathcal{H}_b }\ket{0}_a \propto
\prod_j \rme^{-it  \mathcal{J} (\xi_b-\xi_a)/\xi_a\sigma_x^j \sigma_x^{j+1}} \ket{0}_a$
flips pairs of adjacent spins so that they point in the $+z$-direction. Due to the
qubit-qubit coupling, these local excitations propagate as QPs with velocity $v\approx 2J$ through the chain. 
For an interpretation of the quench dynamics and the time scales indicated in
the plots (all of which scale like $N/J$) in terms of these QPs, see \cite{Viehmann2012}. If $\xi_b$ is in the
paramagnetic phase, the mean free path $l$ of the QPs in a disordered TFIC can be
estimated by $l = 1/(\sigma_\tau \xi_b)^2$ according to the previous section. The
characteristic quasi-$T$-periodic behavior ($T=N/v$) of the local magnetization
after the quench in the non-disordered TFIC can be understood
as a revival of coherence each time QPs initially generated at the
same spot meet again \cite{Rieger2011,Viehmann2012}. This happens when the QPs have
travelled multiples of the chain length $N$. If there should be a significant
probability that two contiguously generated QPs meet again at least once before being
scattered and thus decrease the local magnetization at $t=T$,
the mean free path has to be sufficiently large, $l>2N$. The appearance of significant end-to-end
correlations after the quench (that are stronger than those for $t\rightarrow \infty$) 
requires that QPs generated in the middle of the chain reach the edges of the
chain without being scattered, hence $l > N$. We have performed numerical experiments
which indeed suggest that the corresponding values of $\sigma_\tau$ mark the
transition to a degree of disorder where the described phenomena are no longer
present. In that sense, the distinctive features
of the quench dynamics of the end-to-end correlator are less sensitive to disorder
than those of the local magnetization (and, due to the shorter time-scale, less
sensitive to decoherence or decay). We finally note that also here the effective chain
length can be adjusted by strongly detuning individual qubits
(this can also be used to create local quantum quenches by
`joining' two initially independent chains) and arbitrary effective potentials $\Omega_j$ can be
chosen.

\section{Conclusion}
In the quest for controllable large-scale quantum systems, the framework of circuit
QED offers several advantages, such as fast, high-fidelity readout, great flexibility
in design, and steadily increasing coherence times. However, a potential significant
disadvantage arises from the hardly avoidable static noise and disorder sources in these man-made
devices. The central result of the present work is that also in this respect, there is reason
to be optimistic: The requirements on the homogeneity
of the system parameters for observing interesting (and predictable) many-body
physics in a circuit QED system are not too high to be achievable with present-day or near-future technology.
This underlines the prospects of circuit QED as a promising platform for implementing
quantum simulations of complex quantum many-body Hamiltonians. In addition, we have
shown that circuit QED quantum simulators could be used to study deliberately the effects of tunable
disorder on quantum many-body dynamics.

\ack
We thank I.\ Siddiqi, R.\ Vijay, A.\ Schmidt, and N.\ Henry for discussions. O.V.
thanks the QNL group at UC Berkeley for their hospitality. Support by NIM and the SFB
631 of the DFG is gratefully acknowledged.

\end{document}